\documentclass[a4paper,11pt]{article}
\pdfoutput=1 

\usepackage{jheppub}

\usepackage{amssymb,amsmath,latexsym,bm,amsfonts}
\usepackage{graphicx}
\usepackage{longtable}
\usepackage{color,xcolor}
\usepackage{indentfirst}
\usepackage{subfigure}

\newcommand{\be}{\begin{equation}}
\newcommand{\ee}{\end{equation}}
\newcommand{\bpm}{\begin{pmatrix}}
\newcommand{\epm}{\end{pmatrix}}

\newcommand{\beqn}{\begin{eqnarray}}
\newcommand{\eeqn}{\end{eqnarray}}

\newcommand{\cD}{\mathcal D}

\newcommand{\cT}{\mathcal T}

\newcommand{\cO}{\mathcal O}
\newcommand{\cR}{\mathcal R}

\newcommand{\cS}{\mathcal S}
\newcommand{\cW}{\mathcal W}
\newcommand{\cQ}{\mathcal Q}

\newcommand{\p}{\partial}

\newcommand{\pp}{{++}}

\newcommand{\mm}{{--}}

\newcommand{\cN}{{\mathcal N}}
\def\a{\alpha}

\def\c{\chi}
\def\d{\delta}
\def\e{\epsilon}

\def\o{\omega}
\def\q{\theta}

\def\s{\sigma}

\def\z{\zeta}

\def\L{\Lambda}

\def\S{\Sigma}

\def\rd{{\rm d}}

\newcommand{\bea}{\begin{eqnarray}}
\newcommand{\eea}{\end{eqnarray}}
\newcommand{\non}{\nonumber}
\newcommand{\ba}{\begin{array}}
\newcommand{\ea}{\end{array}}

\newcommand{\qb}{{\bar{\theta}}}

\newcommand{\bsubeq}{\begin{subequations}}
\newcommand{\esubeq}{\end{subequations}}

\def\ri{{\rm i}}

\newcommand{\ve}{\varepsilon}
\newcommand{\cDB}{{\bar\cD}}
\newcommand{\cQB}{{\bar\cQ}}

\newcommand{\pa}{\partial}
\newcommand{\hf}{\frac12}

\newcommand {\cF}{{\cal F}}

\newcommand {\cM}{{\cal M}}
\newcommand {\cU}{{\cal U}}

\newcommand {\cX}{{\cal X}}
\newcommand {\cY}{{\cal Y}}

\title{ \boldmath $T\bar{T}$ deformations with $\mathcal{N}=(0,2)$ supersymmetry}

\author[a]{Hongliang Jiang,}
\author[b]{Alessandro Sfondrini,}
\author[a]{Gabriele Tartaglino-Mazzucchelli}

\affiliation[a]{Albert Einstein Center for Fundamental Physics,
Institute for Theoretical Physics,\\
University of Bern,
Sidlerstrasse 5, CH-3012 Bern, Switzerland}
\affiliation[b]{Institut f\"ur theoretische Physik, ETH Z\"urich\\ Wolfgang-Pauli-Stra{\ss}e 27, 8093 Z\"urich, Switzerland}

\emailAdd{jiang@itp.unibe.ch}
\emailAdd{sfondria@itp.phys.ethz.ch}
\emailAdd{gtm@itp.unibe.ch}

\abstract{%
We investigate the behaviour of two-dimensional quantum field theories with $\mathcal{N}=(0,2)$ supersymmetry under a 
deformation 
induced by the ``$T\bar{T}$'' composite operator. We show that the deforming operator can be defined by a point-splitting 
regularisation in such a way as to preserve $\mathcal{N}=(0,2)$ supersymmetry. As an example of this construction, we work out the 
deformation of a free $\mathcal{N}=(0,2)$ theory, compare to that induced by the Noether stress-energy tensor
  and argue that, despite their apparent difference, they are equivalent on-shell. 
  Finally, we show that the $\mathcal{N}=(0,2)$ supersymmetric deformed action actually possesses  $\mathcal{N}=(2,2)$ symmetry, 
  half of which is non-linearly realised.
  }

\begin{document}
\maketitle
\flushbottom


 
\section{Introduction and summary}

A fruitful way to gain new insight on Quantum Field Theories (QFTs), and indeed on many physical theories, 
is to start by studying a particularly simple theory---for instance one that can be solved completely by virtue of its
 symmetries---and \textit{deform it}. Generally, of course such deformations can be only understood in some perturbative 
 or even formal expansion in a small parameter. In rare cases it is possible to treat the perturbation exactly, at least for 
 the purpose of computing certain observables. Important examples of this type arise for instance for two-dimensional QFTs 
 as \textit{marginal} deformations which preserve the scaling invariance of a two-dimensional relativistic 
 Conformal Field Theory (CFT), and (relevant) \textit{integrable} deformations which introduce a mass scale while preserving 
 an infinite set of symmetries of the original (conformal) QFT. A recent addition to these two classes is that of so-called 
 $T\bar{T}$ deformations of two-dimensional QFTs. They arise by deforming any Poincar\'e-invariant theory by a 
 composite operator built as the determinant of the stress-energy tensor~\cite{Zamolodchikov:2004ce}, leading to 
 an \textit{irrelevant} deformation. Not only does this deformation preserve  many of the symmetries of the underlying 
 theory (which makes it very interesting to study $T\bar{T}$ deformations of integrable QFTs and CFTs) but it modifies 
 the theory's spectrum in a simple, ``solvable'' way~\cite{Smirnov:2016lqw,Cavaglia:2016oda}, whose classical action can be often constructed in closed~form~\cite{Cavaglia:2016oda,Bonelli:2018kik}.%
 \footnote{A number of 
 generalisations of $T\bar{T}$, involving other conserved currents, has also been considered in the 
 literature~\cite{Guica:2017lia, Bzowski:2018pcy, Nakayama:2018ujt,Guica:2019vnb, LeFloch:2019rut}.}

Following these observations, several other interesting properties of such deformations have recently been analysed. 
In the case of integrable theories these deformations can be understood~\cite{Smirnov:2016lqw,Cavaglia:2016oda} as
 a modification of the factorised S~matrix by a universal ``CDD''~\cite{Castillejo:1955ed} factor. 
Such CDD factor is rather peculiar, because it deforms the large-energy \textit{ultraviolet} properties of the 
original theory---in accord with the fact that $T\bar{T}$ deformations are ``irrelevant''---rather than introducing poles as it is 
usually the case for integrable deformations.
Interestingly, the ``$T\bar{T}$'' CDD factor had also appeared in the study of the S~matrix on the worldsheet of 
flat-space strings~\cite{Dubovsky:2012wk,Dubovsky:2012sh, Caselle:2013dra, Cavaglia:2016oda, Chen:2018keo, Baggio:2018gct, Baggio:2018rpv}, 
strongly suggesting that flat-space strings are the $T\bar{T}$ deformation of a \textit{free} theory (a fact that can also be
 substantiated from a Lagrangian or Hamiltonian analysis~\cite{Cavaglia:2016oda,Baggio:2018gct,Baggio:2018rpv}). 
 Indeed $T\bar{T}$ deformations are naturally related~\cite{Baggio:2018gct,Baggio:2018rpv} to the ``uniform'' 
 light-cone gauge which is quite natural for the study of integrable strings 
 theories~\cite{Arutyunov:2004yx,Arutyunov:2005hd, Arutyunov:2006gs}, and the $T\bar{T}$ CDD factor also describes the
  scattering on more general backgrounds such as $AdS_3$ Wess-Zumino-Witten 
  backgrounds~\cite{Baggio:2018gct,Dei:2018mfl, Dei:2018jyj}.%
  \footnote{Such backgrounds are supported by 
  Neveu-Schwarz-Neveu-Schwarz fluxes only; the worldsheet scattering for $AdS_{3}$ backgrounds involving Ramond-Ramond 
  fluxes is substantially more involved~\cite{Borsato:2012ud,Borsato:2014exa, Sfondrini:2014via}, as it may also be understood by 
  worldsheet-CFT considerations~\cite{Berkovits:1999im}, see \textit{e.g.}\ ref.~\cite{Eberhardt:2018exh}.} 
A separate, but equally interesting link between $T\bar{T}$ deformations and $AdS_3$ strings appears in the context of 
holography~\cite{McGough:2016lol,Giveon:2017nie, Giveon:2017myj, Giribet:2017imm, Kraus:2018xrn, Gorbenko:2018oov,Araujo:2018rho, Giveon:2019fgr}, where the deformed two-dimensional theory is \textit{the holographic dual} 
of some ($AdS_3$) gravity or string theory, rather than being a worldsheet theory. 

The ``irrelevant'' nature of $T\bar{T}$ deformations leads to a number of very peculiar consequences. It turns out that 
$T\bar{T}$-deformed QFTs can also be understood as a gravitational theory, and more 
specifically~\cite{Dubovsky:2017cnj,Cardy:2018sdv, Dubovsky:2018bmo,Conti:2018tca} as coupling the original 
two-dimensional theory to Jackiw-Teitelboim gravity~\cite{Jackiw:1984je,Teitelboim:1983ux}. 
For a generic theory, furthermore, the $T\bar{T}$  flow induces a singular behaviour of some of the energy levels for a finite 
value of the deformation parameter in units of the theory's volume~\cite{Smirnov:2016lqw,Cavaglia:2016oda,Aharony:2018bad}. One exception to this scenario are 
supersymmetric QFTs which admit a well-defined flow. This is in good accord with the fact that some superstring theories can be 
described as $T\bar{T}$ deformations; interestingly, this setup seems to be well-defined even when supersymmetry is non-linearly 
realised~\cite{Dei:2018jyj}. Such musings, together with the intrinsic interest in supersymmetric CFTs, spurred a systematic 
investigation of the relation between $T\bar{T}$ deformation and supersymmetry~\cite{Baggio:2018rpv,Chang:2018dge}. So far, 
this investigation focussed on theories with $\mathcal{N}=(0,1)$ and $\mathcal{N}=(1,1)$ supersymmetry.

The aim of this note is to extend such considerations to the case of extended $\mathcal{N}=(0,2)$ supersymmetry. The 
conclusion is that indeed, $T\bar{T}$ deformations can be defined in such a way as to manifestly preserve the extended 
supersymmetry. In particular, the deforming operator can be thought of as the supersymmetric descendant of some suitable 
composite operator. Still, there are a few technical complications and conceptual subtleties with respect to the cases known in the 
literature which we find worth addressing in some detail. 

We begin our note by introducing the $\mathcal{N}=(0,2)$ framework in section~\ref{sec:zerotwo-S-multiplet} 
and~\ref{sec:zerotwo}. 
After reviewing in section~\ref{sec:zerotwo-S-multiplet} 
the structure of the $\cN=(0,2)$ supercurrent multiplet,
in section~\ref{sec:ttbar} we 
construct the $T\bar{T}$ operator as the supersymmetric descendant of a suitably defined composite operator, which is 
constructed out of the coincident-point limit of a quadratic combination of the supercurrents. Here we encounter a new feature: 
the primary operator does not take the form discussed by Smirnov and Zamolodchikov~\cite{Smirnov:2016lqw}, yet it is possible 
to show that the coincident-point limit is free of short-distance singularities.
Interestingly, the well-definedness arguments that we develop in this paper apply also for $T\bar{T}$ deformations of $\cN=(2,2)$ 
supersymmetric QFTs and $J\bar{T}/T\bar{J}$ deformations. Also in these cases, as we will report in the near 
future~\cite{22TTbar,SUSYJTbar},
the primary operators are not of Smirnov-Zamolodchikov type, yet they are well-defined.

As an example of this construction, in section~\ref{sec:deformaction} we work out the supersymmetric $T\bar{T}$ deformation of 
a free $\mathcal{N}=(0,2)$ action, and compare it with the one constructed out of the Noether $T\bar{T}$ operator. The latter is 
most easily obtained from the Green-Schwarz string, see ref.~\cite{Baggio:2018rpv}. We conclude that while apparently different, 
the two deformations are 
 identical on-shell,
giving rise to equivalent flow equations for the spectrum---as expected
from the lower supersymmetric cases~\cite{Baggio:2018rpv}.
Finally, in section~\ref{sect-partial-breaking} we show that the deformation of the free theory also describes the partial 
supersymmetry breaking from $\mathcal{N}=(2,2)$ to $\mathcal{N}=(0,2)$. In fact, such a $T\bar{T}$ deformed action is 
equivalent to the model of partial supersymmetry breaking in two-dimensions by Hughes \& Polchinski  \cite{Hughes:1986dn}, see 
also ref.~\cite{Ivanov:2000nk}, which describes a $\cN=(0,2)$ extension of a 4D Nambu-Goto superstring action. The action also 
shares several analogies with the four dimensional Bagger-Galperin action describing the partial supersymmetry breaking from 
$\mathcal N=2$ to $\mathcal N=1$ in four dimensions \cite{Bagger:1996wp}.  This is a further example of how $
T\bar{T}$-deformed theories possess non-linearly realised (super)symmetries, which is something that would be interesting 
to explore in greater detail. 
Some first results  will appear in the near future~\cite{22TTbar,SUSYDBITTbar}.
In particular, it can be shown that the $T\bar{T}$ deformation of free $\mathcal{N}=(2,2)$ theories 
also describe supersymmetric extensions of a 4D Nambu-Goto superstring action possessing extra $\mathcal{N}=(2,2)$
non-linearly realised supersymmetry. In the $\cN=(2,2)$ case, the resulting actions can in fact be recast in forms that 
are formally identical to the 4D Bagger-Galperin action \cite{Bagger:1996wp} for the supersymmetric extension of the DBI action,
 hinting at a $T\bar{T}$ structure of the latter.
It was already been shown in \cite{Conti:2018tca}, that the classical bosonic 4D DBI action satisfy a peculiar $T\bar{T}$-flow
equation. 
The same property generalise to the supersymmetric case \cite{SUSYDBITTbar}.

\section{Supercurrent multiplet in (0,2)} 
\label{sec:zerotwo-S-multiplet}

Let us  review the structure of the supercurrent
$\cS$-multiplet  of  two dimensional $\mathcal N=(0,2) $ supersymmetric field theories that we will need in our paper.
In this section we follow the paper of  Dumitrescu and Seiberg \cite{Dumitrescu:2011iu}, including their conventions.

In light-cone coordinates, 
a flat 2D $\cN=(0,2)$ superspace is parametrised by 
\begin{equation}
\z^M=(\s^{++},\s^{--},\q^+,\qb^+)\,,
\end{equation}
with $\theta^+$ a complex Grassmann coordinate and $\qb^+$ its complex conjugate.
The spinor covariant derivatives and supercharges are given by
\begin{equation}
\begin{aligned}
\cD_+&=&\frac{\p} {\p\theta^+}-\frac{\ri}{2} \bar \theta^+ \p_{++}
\,, \qquad
\bar \cD_+=-\frac{\p} {\p\bar \theta^+}+\frac{\ri}{2}   \theta^+ \p_{++}
\,,
\\
\cQ_+&=&\frac{\p} {\p\theta^+}+\frac{\ri}{2} \bar \theta^+ \p_{++}
\,, \qquad
\bar \cQ_+=-\frac{\p} {\p\bar \theta^+}-\frac{\ri}{2}   \theta^+ \p_{++}
\,,
\end{aligned}
\end{equation}
and obey the anti-commutation relations
\bea
\{ \cD_+ , \cDB_+ \} =\ri \pa_{++}\,,\qquad
\{ \cQ_+ , \cQB_+ \} =-\ri \pa_{++}\,,
\eea
with all the other (anti-)commutators between $\cD$s, $\cQ$s, and $\pa_{\pm\pm}$ being identically zero.
Given an $\cN=(0,2)$ superfield%
\footnote{For convenience we will equivalently use the notations 
$\cF(\z)=\cF(\s,\q)=\cF(\s^{++},\s^{--},\q^+,\qb^+)$; in particular, we will often indicate collectively by $\q$ the dependence 
on both $\q^+$ and $\qb^+$.}
$\cF(\z)=\cF(\s,\q)$ 
its supersymmetry transformations are given by
\bea
\d_Q \cF
:=
\ri\e^+ \cQ_+ \cF(\s,\q)
-\ri\bar\e^+ \cQB_+ \cF(\s,\q)\,.
\label{susySuperfield}
\eea
Here $\e^+$ and its complex conjugate $\bar\e^+$ are the complex fermionic supersymmetry transformation parameters.
If $F(\s)$ is the operator defined as the $\q^+=\qb^+=0$ component of the superfield 
$\cF(\zeta)$, $F(\s):=\cF(\s,\q)|_{\q=0}$, then 
its supersymmetry transformations are such~that
\begin{equation}
\begin{aligned}
\d_Q F(\s)&=\,
\big[\big(\e^+Q_+-\bar\e^+\bar Q_+\big),F(\s)\big\}
\,,
\\
&=\,
\left[\ri\e^+ \cQ_+ \cF(\s,\q)
-\ri\bar\e^+ \cQB_+ \cF(\s,\q)\right]\big|_{\theta=0}
\,,
\\
&=\,
\left[\ri\e^+ \cD_+ \cF(\s,\q)-\ri\bar\e^+ \cDB_+ \cF(\s,\q)\right]\big|_{\theta=0}
\,.
\end{aligned}
\end{equation}
We will indicate by $Q_+$ and $\bar{Q}_+$ the supersymmetry generator \textit{acting on a component operator}
and distinguish them from $\cQ_+$ and $\cQB_+$, which are linear  differential operators \textit{acting on superfields}.

In two-dimensional $\mathcal N=(0,2)$ supersymmetric field theories, the supercurrent $\cS$-multiplet 
  is defined by the following 
constraints  \cite{Dumitrescu:2011iu}
\bsubeq \label{eq:Smultiplet}
\bea
\p_\mm\cS_\pp{}_{\phantom{+}} 
&=&   \cD_+ \cW_- -\bar\cD_+ \bar \cW_- \,,
\\
\cD_+ \cT_{----} & =&  \frac12 \p_{--} \bar\cW_- \,,
\\
\bar \cD_+ \cT_{----} &=&  \frac12 \p_{--}  \cW_- \,,
\\
\bar\cD_+  \cW_- &=&C
~,
\\
\cD_+  \bar{\cW}_-&=&-\bar{C}
\,,
\eea
\esubeq
where the complex constant $C$ is associated with a space-time brane current. Since this term leads to
symmetry breaking~\cite{Hughes:1986dn,Dumitrescu:2011iu}, in this paper  for simplicity  we will set it to zero, $C=0$. 

In components, the supercurrent $\cS$-multiplet is given by 
\begin{equation}
\begin{aligned}
\cS_{++} &= \phantom{+}j_{++\phantom{++}}
- \ri \theta^+ S_{+++} 
-\ri \bar \theta^+ \bar S_{+++} 
&&-\theta^+\bar \theta^+ T_{++++} 
\,,
\\
\cW_- &=  -\bar S_{+--\phantom{+}} 
-\ri \theta^+ \Big(T_{++--}+\frac{\ri}{2} \p_{--} j_{++} \Big)
&&+\frac{\ri}{2}  \theta^+\bar \theta^+ \p_{++} \bar S_{+--} 
\,,
\\
\cT_{----} &= \phantom{+}T_{----} 
-\frac12 \theta^+ \p_{--} S_{+--} 
+\frac12 \bar \theta^+ \p_{--} \bar S_{+--} 
&&+\frac14   \theta^+\bar \theta^+\p_{--}^2 j_{++}
\,.
\end{aligned}
\end{equation}
The $j_{++}(\s)$, $S_{+\pm\pm}(\s)$, $\bar S_{+\pm\pm}(\s)$, $T_{\pm\pm\pm\pm}(\s)$, and $T_{++--}(\s)$ fields
arise as the lowest, $\q^+=\qb^+=0$, components of the superfields $\cS_{++}(\z)$, $\cW_-(\z)$,
$\bar\cW_-(\z)$, and $\cT_{----}(\z)$
together with their descendants,
\begin{equation}
\label{descendants}
\begin{aligned}
\cS_{+++}(\z)&:=\ri \cD_+\cS_{++}(\z)\,,&
\bar\cS_{+++}(\z)&:=-\ri \cDB_+\cS_{++}(\z)\,,
\\
\cT_{++++}(\z)&:=-\frac{1}{2}[\cD_+,\cDB_+]\cS_{++}(\z)\,,\qquad&
\cT_{++--}(\z)&:=\frac{\ri}{2}(\cD_+\cW_-(\z)+\cDB_+\bar\cW_-(\z))\,.
\end{aligned}
\end{equation}
In the paper we will also use the definitions
\be
\cT(\z):= \cT_{++--}(\z)\equiv\cT_{--++}(\z)
\,,\qquad
\Theta(\s):=\cT|_{\q=0}=T_{++--}(\s)\equiv T_{--++}(\s)
\,.
\ee
From the supercurrent equations  \eqref{eq:Smultiplet} together with the definitions \eqref{descendants}, 
one can derive the conservation
equations
\begin{equation}
\label{conserT}
\begin{aligned}
& \p_{++} \cS_{+--}(\z) &=&\, - \p_{--} \cS_{+++}(\z)\,,\\
& \p_{++} \cT_{----}(\z) &=&\, -\p_{--} \cT(\z) \,,\\
& \p_{++} \cT(\z) &=&\, - \p_{--} \cT_{++++}(\z) \,.
\end{aligned}
\end{equation}
These imply that the supersymmetry currents
$S_{+\pm\pm}$ and $\bar S_{+\pm\pm}$  are conserved
while the energy-momentum tensor $T_{\mu\nu}$ is real, symmetric and conserved. 

It is possible to modify the $\cS$-multiplet by a class of ``improvement terms'' without changing its defining constraint 
equations \eqref{eq:Smultiplet}. This is analogous to how the energy-momentum tensor can be modified 
by  improvement terms that do not affect its conservation equations.
The improvement transformations 
that leave invariant the $\cS$-multiplet conservation equations are
\begin{equation}\label{improvement}
\begin{aligned}
&\cS_{++\phantom{++}} \quad\rightarrow\quad  \tilde \cS_{++\phantom{++}}
=\cS_{++\phantom{++}}+2\,[\cD_+, \bar \cD_+]\, \cU\,, \\
&\cW_{-\phantom{++}}\, \quad\rightarrow\quad  \tilde \cW_{-\phantom{++}}\,=\cW_{-\phantom{++}}\,+ 2\p_{--} \bar \cD_+  \cU\,,\\
&\cT_{----} \quad\rightarrow\quad  \tilde\cT_{----}=\cT_{----}+\p_{--}^2  \cU\,,
\end{aligned}
\end{equation}
where $\cU(\z)$ is a real scalar superfield with lowest component field $U(\s):=\cU(\z)|_{\q=0}$. 
The improvement transformations of the energy-momentum tensor 
induced by \eqref{improvement} are
\begin{equation}\label{components-improv}
\begin{aligned}
&T_{++++} \quad\rightarrow\quad \tilde T_{++++}\,=\, T_{++++}+ \p_{++}^2  U\,,\\
&T_{----} \quad\rightarrow\quad \tilde T_{----}\,=\, T_{----}+ \p_{--}^2  U\,,\\
&T_{++--} \quad\rightarrow\quad \tilde T_{++--}\,=\, T_{++--}- \p_{++}\p_{--}   U\,.
\end{aligned}
\end{equation}
It is clear that $\tilde{T}_{\mu\nu}$ is also real, symmetric and conserved.

The $\cS$-multiplet described above is
 the most general supercurrent multiplet for a Lorentz invariant and $\cN=(0,2)$
supersymmetric quantum field theory in two space-time dimensions. 
In some cases, the multiplet is decomposable and the currents can be improved. 
A case that will play a central role in our paper is the $\cR$-multiplet.
This arises when $C=0$, which is indeed the case under our consideration, 
and when there is a well-defined real superfield $\cR_{--}(\z)$ 
resolving the chirality constraint of $\cW_-(\z)$ as
\be
\label{eq:W-R-relation}
\cW_- =\ri \bar\cD_+ \cR_{--}
~.
\ee
The defining conservation equations for the $\cR$-multiplet  can then be  written as%
\footnote{The $\cR$-multiplet 
conservation equations are also derived  from supergravity in the appendix \ref{sugra-supercurrent}.  }
\bsubeq
\label{Rmultiplet}
\beqn
\p_{--} \cR_{++}+ \p_{++}\cR_{--} &=&0
\, ,
\\
\cD_+ \Big(   \cT_{----}  +\frac{\ri}{2} \p_{--} \cR_{--} \Big) &=&0
\, ,
\\
\bar \cD_+ \Big(   \cT_{----}  -\frac{\ri}{2} \p_{--} \cR_{--} \Big) &=&0
\,,
\eeqn
\esubeq
where $\cR_{++}\equiv \cS_{++}$.  
The main consequence of the extra constraints imposed on the $\cR$-multiplet is
the existence of an extra conserved vector current $j_{\pm\pm}(\s)$
\be
\p_{++} j_{--} +\p_{--} j_{++}=0
~,
\ee
with $j_{--}:=\cR_{--}|_{\q=0}$.
This current is associated to 2D $\cN=(0,2)$ theories
possessing  a $U(1)_R$ $R$-symmetry.
As described in appendix \ref{sugra-supercurrent},  the $\cR$-multiplet naturally arises from $\cN=(0,2)$ Poincar\'e 
supergravity. In the explicit examples of 2D $\cN=(0,2)$ theories which we will consider in our paper, we will always compute
the supercurrent multiplet by means of coupling the theory to supergravity with a procedure that mimics the calculation 
of the Hilbert stress-energy tensor from gravity. This approach will guarantee that the resulting supercurrent multiplet will be 
an $\cR$-multiplet.

We conclude this section by mentioning that in a 2D $\cN=(0,2)$ superconformal field theory (SCFT) the 
$\cS$-multiplet can be further simplified. In fact, for a SCFT with $\cN=(0,2)$ supersymmetry,
$C=0$ and $\cW_-$ can be set to zero by an improvement transformation. 
Then the $\cS$-multiplet only contains a right-moving superfield current $\cS_{++}(\z)$, $\p_{--} \cS_{++}=0$, 
and a left-moving anti-chiral superfield $\cT_{----}(\z)$, $\cD_+ \cT_{----}(\z) =0$. This leads to a set of left- and right-moving 
currents in components.

\section{The \texorpdfstring{$T\bar T$}{TTbar} operator and \texorpdfstring{$\cN=(0,2)$}{N=(0,2} supersymmetry} 
\label{sec:zerotwo}

After having described in the previous section the structure of the $\cS$-multiplet, we are 
ready to prove that the $T\bar{T}$ operator \cite{Zamolodchikov:2004ce}
\bea
O(\s)=T_{++++}(\s)\,T_{----}(\s)-\big[\Theta(\s)\big]^2
~,
\label{component-TTbar}
\eea
is a supersymmetric descendant, in complete analogy to the $\cN=(0,1)$ and $\cN=(1,1)$ cases 
first studied in \cite{Baggio:2018rpv,Chang:2018dge}.

\subsection{The \texorpdfstring{$T\bar T$}{TTbar} primary operator}
\label{sec:ttbar}

We propose the $\cN=(0,2)$ supersymmetric primary $T\bar T$  
operator to be given by the following combination of the $\cS$-multiplet superfields
\be
\cO_{--}(\z):=\cT_{----}(\z)\cS_\pp(\z)   -\bar \cW_-(\z) \cW_- (\z)
~.
\label{primary-TTbar}
\ee
In fact, it is a straightforward exercise to show that the following relation holds
\bea
\cD_+\cDB_+\cO_{--}(\z)
&=&
-\cT_{----}(\z)\cT_{\pp\pp}(\z)  
+[\cT(\z)]^2
\non\\
&&
+\pa_\mm \Big[
\frac{1}{4}\cS_\pp(\z)\pa_{--}\cS_{++}(\z)
-\hf\cW_-(\z)\cD_{+}\cS_\pp(\z)  
+\hf\bar\cW_-(\z)\cDB_+\cS_\pp(\z)
\Big]
\non\\
&&
+\pa_\pp\Big[
\frac{\ri}{2}\cT_{----}(\z)\cS_\pp(\z) 
-\frac{\ri}{2}\bar \cW_-(\z) \cW_- (\z)
\Big]
\non\\
&&
+\,\textrm{EOM's}
~,
\label{DDO}
\eea
where with ``EOM's'' we mean terms that are identically zero when the $\cS$-multiplet 
conservation equations \eqref{eq:Smultiplet} are used.
Since the conservation equations classically hold only when the equations of motion are satisfied, the previous results show
that, on-shell and up to total derivatives, the previous descendant is equivalent to the $T\bar{T}$ operator
\eqref{component-TTbar}. Quantum mechanically, the same statement is true for the corresponding operators since 
conservation equations (Ward identities) hold in correlation functions (up to contact terms).

If we now define
\begin{equation}
O_{--}(\s)\equiv \cO_{--}(\z)\big|_{\q=0}
\,,
\end{equation}
by using the previous results, up to total derivatives and EOM's, 
the operator $O(\s)$, eq.~\eqref{component-TTbar},
satisfies
\bea
O(\s)
=\int \rd\q^+\rd\qb^+ \,\cO_{--}(\z)
=-\cD_+\cDB_+\cO_{--}(\z)\big|_{\q=0}
=-\big\{Q_+,\big[\bar Q_+,O_{--}(\s)\big]\big\}
~.
\label{OQQO}
\eea
Then, $O_{--}(\s)$ is the supersymmetric primary operator of the multiplet containing $O(\s)$ as its bottom component.
Hence the $T\bar{T}$ deformation for an $\cN=(0,2)$ supersymmetric quantum field theory
is manifestly supersymmetric since eq.~\eqref{OQQO} implies
\bea
{\Big[}Q_+\,,\,\int\rd^2\s\, O(\s){\Big]}
=
{\Big[}\bar Q_+\,,\,\int\rd^2\s\, O(\s){\Big]}
= 0
~.
\eea

The $T\bar T$ primary operator \eqref{primary-TTbar} is   defined uniquely by the requirement that $O(\s)$ is its descendant,
up to conservation equations and total derivatives. Another virtue enjoyed by $\cO_{--}(\z)$
is that its form, up to total derivatives,  is invariant under the improvement transformation \eqref{improvement}:
 \be
 \mathcal O_{--}
 \rightarrow 
 \mathcal {\tilde O}_{--}= \tilde\cS_\pp\tilde \cT_{----}  - \tilde{\bar\cW}_-\tilde \cW_- = \mathcal O_{--}+\text{total derivatives}
 ~.
 \ee
 This is not too surprising since the combination $(T_{----}T_{++++}-\Theta^2)$ is invariant under the improvement 
 transformations \eqref{components-improv}.

 \subsection{Point-splitting and well-definedness}
 \label{point-plitting-section}

As shown by Zamolodchikov in his seminal work \cite{Zamolodchikov:2004ce},
one of the main properties of the $T\bar T$ operator $O(\s)$ is to be free of short-distance divergences
and hence to be a well-defined, though irrelevant, composite local operator. 
More in general, it was later shown by Smirnov and Zamolodchikov in \cite{Smirnov:2016lqw}
that given two pairs of conserved currents $(A_s,B_{s+2})$ and $(A'_{s'},B'_{s'-2})$ such that 
\bea
\partial_{++} A_{s} &= - \partial_{--} B_{s+2}~,\qquad
\partial_{--} A'_{s'} &= - \partial_{++} B'_{s'-2}~,
\eea
then the bilocal operator $[A_{s}(\sigma)\, A'_{s'}(\sigma')-B_{s+2}(\sigma)\, B_{s'-2}(\sigma')]$
is free of short-distance divergences and, up to total derivative terms,
independent of the separation $(\sigma - \sigma')$.  Here $s$ and $s'$ label spins.
Hence, any composite local operator of the ``Smirnov-Zamolodchikov'' type,
\begin{equation}
\int\rd^2\s\Big[A_{s}(\s)\, A'_{s'}(\s)-B_{s+2}(\s)\, B_{s'-2}(\s)\Big]\,,
\end{equation}
is well-defined. This is indeed the case of the bosonic $T\bar{T}$ operator $O(\s)$.
In the case of $\cN = (0,1)$ and $\cN = (1, 1)$ supersymmetric $T\bar{T}$ deformations, 
the primary $T\bar{T}$ superfield operators are still
of Smirnov-Zamolodchikov type \cite{Baggio:2018rpv,Chang:2018dge}, 
hence well-defined and leading  to a whole multiplet of well-defined composite operators.

In the $\cN=(0,2)$ case, the $T\bar{T}$ primary operator $O_{--}(\s)=\cO_{--}(\z)|_{\q=0}$
from eq.~\eqref{primary-TTbar} is not of Smirnov-Zamolodchikov type.
It is then natural to wonder whether the primary $O_{--}(\s)$ can also be defined by a similar point-splitting procedure 
without incurring in short-distance singularities. Were this not the case, we would have an apparent clash between 
supersymmetry and the 
structure of the $T\bar{T}$ deformation at the quantum level. We shall see below that, owing to supersymmetry, 
$\cO_{--}(\s,\q)$ can be indeed defined in superspace by 
a point-splitting procedure \textit{of the bosonic coordinate}%
\footnote{As for the $\theta$s, we can straightforwardly set them to be 
equal since no divergence of the form \textit{e.g.} $1/(\theta-\theta')$ arises for the Grassmann coordinates.
See section 6 of \cite{Gates:1983nr} for examples of point-splitting techniques in superspace.}
$\s$ in analogy with the arguments 
of~\cite{Zamolodchikov:2004ce,Smirnov:2016lqw}.   
 
 Let us consider  a point-split version of the $\cN=(0,2)$ primary $T\bar{T}$ operator,
 \bea
 \cO_{--}(\s,\s',\q)
 :=
 \cT_{----}(\s,\q)\,\cS_\pp(\s',\q)   
 -\bar \cW_-(\s,\q)\, \cW_-(\s',\q)
 ~.~~~~~~
 \label{pointsplitted-primary}
 \eea
We want to show that the previous bilocal superfield is free of short distance divergences
in the limit $\s\to\s'$. 
Following refs.~\cite{Zamolodchikov:2004ce,Smirnov:2016lqw}, let us compute 
$\pa_{\pm\pm} \cO_{--}(\s,\s',\q)$. We start by defining
\begin{equation}
\cO_{--}(\z,\z') := [\cT_{----}(\z)\,\cS_\pp(\z')    -\bar \cW_-(\z)\, \cW_-(\z')]\,,
\end{equation}
which is the fully superspace point-split version of 
$\cO_{--}(\z)$---from that, we will easily extract $ \cO_{--}(\s,\s',\q)=\cO_{--}(\z,\z')|_{\q=\q'}$.
Let us compute $\pa_{\pm\pm} \cO_{--}(\z,\z')$.
After some straightforward algebraic manipulation, and by using the fact that $\pa_{++}=-\ri(\cD_+\cDB_++\cDB_+\cD_+)$,
it is possible to derive the following result
\bea
\pa_{\pm\pm}\cO_{--}(\z,\z')
&=&
0
+\,\textrm{EOM's}
+(\pa+\pa')[\cdots]
+(\cD+\cD')[\cdots]
~.
\label{pammO-0_pappO-00}
\eea
Here with ``EOM's'' we again refer to terms that are identically zero once the conservation equations
\eqref{eq:Smultiplet} for the $\cS$-multiplet are used while with the last two terms
in \eqref{pammO-0_pappO-00} we  indicate terms that are
superspace total derivatives, such as for example 
the vector derivatives $(\pa_{\pm\pm}+\pa'_{\pm\pm})$ 
or the spinor derivatives $(\cD_++\cD'_+)$, acting on bilocal operators.
The precise expressions for \eqref{pammO-0_pappO-00} are given by 
eq.~\eqref{pammO} and \eqref{pappO} in appendix \ref{appendix-for-point-splitting-section}
where we collect some technical results and explanations  that support the analysis of this subsection.
When we consider the coincident limit  $\q=\q'$ in the Grassmann coordinates,
thanks to \eqref{DQ},
equation \eqref{pammO-0_pappO-00} can be rewritten in the following useful form
\bea
\pa_{\pm\pm}\cO_{--}(\s,\s',\q)
&=&
0
+\Big\{
\,\textrm{EOM's}
+(\pa+\pa')[\cdots]
+(\cQ+\cQ')[\cdots]
\Big\}\Big|_{\q=\q'}
~,
\label{pammO-0_pappO-0}
\eea
where the supersymmetry generators appear instead of the covariant spinor derivatives.
In the previous expression $(\pa+\pa')$ generates  translations in the $\s$ and $\s'$ coordinates
while schematically $(\cQ+\cQ')$ generates supersymmetry transformations
of the bilocal operators they act upon.%
\footnote{Note that, thanks to the super-Leibniz rule satisfied by 
$\frac{\pa}{\pa\q^+}$ and $\frac{\pa}{\pa\qb^+}$,
the operation of taking the $\q=\q'$ limit and acting on bilocal superfields
with Grassmann dependent differential operator,
such as $(\cD+\cD')$ or $(\cQ+\cQ')$, commute. Hence \eqref{pammO-0_pappO-0} is well defined.
See appendix \ref{appendix-for-point-splitting-section} for more comments on this point.}
The results presented above are reminiscent of Zamolodchikov's argument to prove the well-definedness
 of the bosonic $T\bar{T}$ operator, as well as to the arguments used in the $\cN=(0,1)$ and $\cN=(1,1)$ 
cases \cite{Baggio:2018rpv,Chang:2018dge}. A new feature with respect to those cases are the supersymmetry-transformation 
terms, which represent a natural generalisation of the translation contribution. 
Still, Zamolodchikov's OPE argument of~\cite{Zamolodchikov:2004ce} can be used almost identically here
in the $\q=\q'$ limit, which is sufficient to probe the short distance singularities in $\s\to\s'$.  Let us briefly review it.

By setting to zero the EOM's terms,
the left hand side of \eqref{pammO-0_pappO-0} has an OPE expansion of the form
\bea
\sum_I\pa_{\pm\pm}F_I(\s-\s')\cO_I(\s',\q)
~,
\label{lhs}
\eea
with $\{\cO_I(\z')\}$ a complete set of local superfield operators, depending on the Grassmann coordinate $\theta'=\theta$.
Similarly, the right hand side of  \eqref{pammO-0_pappO-0} will schematically be of the form
\bea
\sum_IA_I(\s-\s')\,\cQ'\cO_I(\s',\q )
+\sum_IB_I(\s-\s')\,\pa'\cO_I(\s',\q )
~,
\eea
which is equivalent to
\bea
\sum_IA_I(\s-\s')\,\cD'\cO_I(\s',\q )
+\sum_IC_I(\s-\s')\,\pa'\cO_I(\s',\q )
~.
\eea
Hence the OPE of $\partial_{\pm\pm}\mathcal{O}_{--}(\sigma,\sigma',\theta)$  involves only derivatives and supercovariant 
derivatives of local operators. This means that the OPE of $\mathcal{O}_{--}$ involves only such derivatives, or terms 
$F_I(\sigma-\sigma') \mathcal{O}_I(\sigma' ,\theta )$ such that the coefficients $F_I$ are actually \textit{constant} 
(so that $\partial_{\pm\pm}F_I=0$), \textit{i.e.}\ \textit{regular terms}. 
Then, the point-splitted superfield operator leads to the definition of the composite $\cN=(0,2)$ $T\bar{T}$ 
primary:
\be
\cO_{--}(\s,\s',\q')
=
\cO_{--}(\z')
\,+\,
{\rm derivative~terms}~,
\ee
arising from the regular, non-derivative  part of the OPE---precisely as for the purely bosonic $T\bar{T}$ operator 
of \cite{Zamolodchikov:2004ce}.
When considering the integral of $\mathcal{O}_{--}(\z)$  in superspace,
only the regular terms  in the OPE would contribute. 
As a result the integrated operator
\bea
S_{\cO} = 
\int\rd^2\s\,\rd\q^+\rd\qb^+\,\lim_{\ve\to0}\cO_{--}(\sigma, \sigma+\ve, \theta) 
=\int\rd^2\s\,\rd\q^+\rd\qb^+\,:\cO_{--}(\sigma, \sigma, \theta):
\eea
is free of any short distance divergence and well-defined.

As a further evidence of the consistency of the previous point-splitting argument with supersymmetry, 
one can consider the point-split version of eq.~\eqref{DDO} and show that
\bea
(\cD_++\cD'_+)(\cDB_++\cDB'_+)\cO_{--}(\z,\z')
&=&
-\cT_{----}(\z)\cT_{++++}(\z')
+\cT(\z)\cT(\z')
\non\\
&&
+\textrm{EOM's}
+(\pa+\pa')\left[\cdots\right]
~,
\eea
with the terms in the ellipsis  being a simple point-split generalisation of the total derivatives
appearing in eq.~\eqref{DDO}. This shows explicitly that the descendant of the point-split
primary $T\bar{T}$ operator is equivalent, up to Ward identities and total vector derivatives, to
the point-split version of the descendant (standard) $T\bar{T}$ operator.

 \section{Deforming the free supersymmetric action}\label{sec:deformaction}

After having described some general properties of the $\cN=(0,2)$ $T\bar{T}$ operator, 
we are ready to study $T\bar{T}$ deformations. We will focus our attention for the rest of the paper on the simplest possible case:
the $T\overline{T}$ deformation of a free action with $\mathcal{N}=(0,2)$ supersymmetry. Though simple, we will see that
a detailed analysis of this model is nontrivial and rich.%
\footnote{Note that in our paper the definition of the $T\bar{T}$ flow is purely 
field theoretical and follows in spirit the original prescription of Zamolodchikov \cite{Zamolodchikov:2004ce}.
Alternative descriptions based on the relation with two-dimensional 
gravitational theories were pursued in~\cite{Dubovsky:2017cnj,Cardy:2018sdv, Dubovsky:2018bmo,Conti:2018tca}.
It would be very interesting to extend these results to the supersymmetric case. Chiral supersymmetric 
theories, as $\cN=(0,2)$, once coupled to supergravity are typically plagued by gravitational anomalies and
it would be important to understand the role of the anomalies in $T\bar T$-deformation interpreted in terms of 2D quantum 
gravity.  This is also an important issue for some non-supersymmetric $T\bar T$-deformations, 
like the deformation of systems of chiral fermions.}

Before turning to the supersymmetric analysis,
let us briefly recall the form of the $T\overline{T}$ deformation of a \textit{bosonic} action for a complex boson whose 
 free action is
\begin{equation}
 S_{0,\text{bos}}=\frac14 \int \rd^2\sigma \Big[ \p_{++} \phi \p_{--}\bar\phi  + \p_{++} \bar \phi \p_{--} \phi     \Big]\,.
 \label{freebos}
\end{equation}
The aim of our analysis is to consider the supersymmetric extension of this simple model and derive its integrated
deformation.
 The $T\bar T$-deformed action of the above free scalar can be compactly written as~\cite{Cavaglia:2016oda}
\begin{equation}
\label{eq:bosonTTbar}
 S_{\alpha,\text{bos}}=  \int \rd^2\sigma   \frac{\sqrt{1+2 \alpha  x +\alpha^2 y^2}-1}{4\alpha}\,,
\end{equation}
 where we have introduced the short-hand notation
\begin{equation}
  \label{eq:xydef}
 x=\p_\pp \phi \p_\mm \bar \phi +\p_\pp  \bar \phi \p_\mm  \phi , \qquad 
y= \p_\pp \phi \p_\mm \bar \phi -\p_\pp  \bar \phi \p_\mm  \phi  \,.
\end{equation}
It is such that
\bea
\frac{\pa S_{\alpha,\text{bos}}}{\pa\a}
=-
\hf  \int \rd^2\sigma  \,O(\s)
~,
\eea
where   $O(\s)=\det[T_{\mu\nu}(\alpha)]$ with the stress-energy tensor of \eqref{eq:bosonTTbar} given by
\begin{equation}
\label{eq:bosTmunu}
 T_{\pm\pm,\pm\pm}=- \frac{  \p_{\pm\pm}\phi   \p_{\pm\pm} \bar\phi }{ \sqrt{1+2 \alpha  x +\alpha^2  y^2}}\,, \qquad
  T_{\pm\pm,\mp\mp}=\frac{ 1+\alpha x -\sqrt{1+2 \alpha  x +\alpha^2  y^2} }{ 2\alpha\sqrt{1+2 \alpha  x +\alpha^2  y^2}}\,.
\end{equation}

It is not difficult to make an educated guess for the $\mathcal{N}=(0,2)$ supersymmetric extension of such a bosonic action, 
by requiring firstly the action is manifestly supersymmetric and secondly that its bosonic part is given by 
eq.~\eqref{eq:bosonTTbar}. We can easily take care of the former requirement by working in superspace; as for the latter, 
let us note that the bosonic action can be recast in the form
\begin{equation}
 \label{eq:bosonTTbar2}
 S_{\alpha,\text{bos}}=
 -  \int \rd^2\sigma \Big(- \frac{  x}{4 } + \alpha \frac{ \p_\pp \phi \p_\mm\phi \p_\pp \bar \phi \p_\mm \bar\phi }
 {  1+\alpha x+\sqrt{1+2 \alpha  x +\alpha^2 y^2}  }  \Big)\,.
\end{equation}
 This immediately suggests the following manifestly off-shell supersymmetric action:
\begin{equation}
 \label{eq:susyTTbaraction}
 S_{\alpha}=
 -\int \rd^2\sigma\, \rd\theta^+\rd\bar\theta^+   \Big(- \frac{\ri}{2 } \bar\Phi \p_{--} \Phi + \alpha 
 \frac{ \cD_+ \Phi  \bar\cD_+ \bar \Phi \p_\mm\Phi   \p_\mm \bar\Phi }
 {  1+\alpha  \mathcal X+\sqrt{1+2 \alpha   \mathcal X +\alpha^2 \mathcal Y^2}  }  \Big)\,,
\end{equation}
which we have written in terms of the chiral and anti-chiral superfields $\Phi$ and $\bar{\Phi}$. They satisfy 
$\cD_+ \bar\Phi=\bar\cD_+\Phi=0$ and are given by the following expansion in component fields
\begin{equation}
\label{eq:superfields}
 \Phi=\phi +  \theta^+ \psi_+ - \frac{\ri}{2} \theta ^+ \bar \theta^+ \p_\pp\phi, \qquad
\bar \Phi=\bar\phi - \bar \theta^+ \bar\psi_+ + \frac{\ri}{2} \theta ^+ \bar \theta^+ \p_\pp \bar \phi\,.
\end{equation}
We have also introduced the bilinear combinations
\begin{equation}
\mathcal X=\p_\pp \Phi \p_\mm \bar \Phi +\p_\pp  \bar \Phi \p_\mm  \Phi\, , \qquad 
\mathcal Y= \p_\pp \Phi \p_\mm \bar \Phi -\p_\pp  \bar \Phi \p_\mm  \Phi   \,.
\end{equation}
Notice that they are simply related to the short-hand $x,y$ of eq.~\eqref{eq:xydef} as
\begin{equation}
 x=\mathcal X|_{\theta=0}, \qquad  y=\mathcal Y|_{\theta=0}\,.
\end{equation}

Actually, there is a very natural reason why the previous educated guess should work. 
The action \eqref{eq:bosonTTbar2} describes the Nambu-Goto action for a four dimensional string in a uniform light-cone  gauge.
Its $\cN=(0,2)$ superstring extension has been studied long ago by Hughes and Polchinski
 in one of the seminal works on partial 
supersymmetry breaking \cite{Hughes:1986dn}. In fact, their results lead to an action equivalent to 
\eqref{eq:susyTTbaraction} which in turn admit a nonlinearly realised extra $\cN=(2,0)$ supersymmetry.
See ref.~\cite{Ivanov:2000nk} for a more recent analysis, that we will follow quite closely in section 
\ref{sect-partial-breaking} when we will review and extend the results on partial supersymmetry breaking of
\eqref{eq:susyTTbaraction}.
Considering that the action \eqref{eq:susyTTbaraction} was already known to be a $\cN=(0,2)$ extension of 
\eqref{eq:bosonTTbar2}, it is absolutely natural to guess that it describes the supersymmetric $T\bar{T}$ flow.
Let us now validate this guess.

\subsection{Some limits of the deformed action}
As a first sanity check of our proposal we consider it in some limits, starting from $\alpha\to0$. In that case, 
it is manifest that only the first summand in~\eqref{eq:susyTTbaraction} survives, so that we find
\bea
 S_0\,&=&\frac{\ri}{2} \int \rd^2\sigma\, \rd\theta^+\rd\bar\theta^+ \; \bar\Phi \p_{--} \Phi
 \non\\
 &=&
 \frac14\int \rd^2 \sigma \Big( \p_{++} \phi \p_{--}\bar\phi  + \p_{++} \bar \phi \p_{--} \phi +2\ri \bar\psi_+\p_\mm \psi_+ \Big)\,,
   \label{eq:freeaction02}
   \eea
which is indeed the free action for the supersymmetric extension of \eqref{freebos}.

Furthermore, we can check which form the action~\eqref{eq:susyTTbaraction} takes when setting some of its fields to zero.
 To this end, it is useful to note that
\begin{equation}
 \cD_+ \Phi  =\psi_+ 
 -\ri\bar \theta^+ \p_\pp \phi 
 +\frac{\ri}{2}\theta^+\bar\theta^+ \p_\pp \psi_+\,,
  \qquad  \bar\cD_+\bar \Phi  = \bar\psi_+ 
  +\ri  \theta^+ \p_\pp \bar\phi
  -\frac{\ri}{2}\theta^+\bar\theta^+ \p_\pp \bar\psi_+\,.
\end{equation}
Setting now all $\psi=\bar{\psi}=0$ we find that the action takes the form
\bea
 S_{\alpha,\text{bos}} 
 &=& 
 -\int \rd^2\sigma\, \rd\theta^+\rd\bar\theta^+   \Big(- \frac{\ri}{2 } \bar\Phi \p_{--} \Phi + \alpha 
 \frac{( -\ri\bar \theta^+ \p_\pp \phi )(\ri  \theta^+ \p_\pp \bar\phi) \p_\mm\Phi   \p_\mm \bar\Phi }
 {  1+\alpha  \mathcal X+\sqrt{1+2 \alpha   \mathcal X +\alpha^2 \mathcal Y^2}  }  \Big)
  \non\\
  &=&   \int \rd^2\sigma       \Big(  \frac{   1}{4 } x- \alpha 
 \frac{   \p_\pp \phi   \p_\pp \bar\phi \p_\mm\phi  \p_\mm \bar\phi}
 {  1+\alpha  x+\sqrt{1+2 \alpha  x +\alpha^2y^2}  }  \Big)\,,
\eea
as expected. Conversely, keeping track of the fermions but setting the bosons to zero, $\phi=\bar{\phi}=0$, we have
\bea
 S_{\alpha,\text{ferm}}&=& -  \int \rd^2\sigma\, \rd\theta^+\rd\bar\theta^+ \Big(- \frac{\ri}{2 } \bar\Phi \p_{--} \Phi + \alpha 
\psi_+   \bar\psi_+     (\theta^+ \p_\mm\psi_+ )( - \bar\theta^+ \p_\mm \bar\psi_+) 
 \Big)
 \non\\
& =&
\int \rd^2\sigma\; \Big( \frac{\ri}{2}  \bar\psi_+\p_\mm \psi_++ \alpha  \psi_+ \bar\psi_+\p_\mm \psi_+  \p_\mm \bar\psi_+ \Big)\,.
\eea
 This is indeed the $T\bar{T}$-deformation of a complex free-fermion action.

\subsection{Constructing the deforming operator}

If the action \eqref{eq:susyTTbaraction} satisfies the supersymmetric $T\bar{T}$ flow equation, 
then the following equation must be satisfied
\begin{equation}
\label{eq:TTbarflowexplicit}
\p_\alpha S_{\alpha}=-\frac12\int \rd^2\sigma\,   \rd\theta^+\rd\bar\theta^+    \Big(  \cS_\pp \cT_{----}  - \bar \cW_- \cW_-  \Big)\,.
\end{equation}
It is easy to compute the left-hand side of this equation,
\begin{equation}
\p_\alpha S_{\alpha}= 
-   \int \rd^2\sigma \rd\theta^+ \rd\bar\theta^+   
 \frac{ \cD_+ \Phi  \bar\cD_+ \bar \Phi \p_\mm\Phi   \p_\mm \bar\Phi }
 {  1+\alpha  \mathcal X+\sqrt{1+2 \alpha   \mathcal X +\alpha^2 \mathcal Y^2}  }
  \frac{1}{\sqrt{1+2 \alpha   \mathcal X +\alpha^2 \mathcal Y^2}}\,.
\end{equation}
As for the right-hand side of eq.~\eqref{eq:TTbarflowexplicit}, we can find the supercurrents by coupling the theory to 
supergravity---in analogy to how the Hilbert stress energy tensor is computed by coupling the theory to a metric. 
For this task, we can use off-shell supergravity techniques developed in the 1980s,
see appendix \ref{sugra-supercurrent} for detail and references. 
With this analysis at hand, it is straightforward, though lengthy, to derive the supercurrent $\cR$-multiplet 
of the action \eqref{eq:susyTTbaraction}.
The details of the supercurrent computation, that might be in principle used in the future also for more complicated models,
are relegated to appendix \ref{sugra-supercurrent}. The results of our analysis are as follow.
We find
\begin{equation}
\begin{aligned}
  \cS_\pp= \cR_\pp&=-\frac{  \cD_+ \Phi  \bar\cD_+ \bar \Phi }{\sqrt{1+2 \alpha   \mathcal X +\alpha^2 \mathcal Y^2}}\,,
   \\
\cR_{--}&= \frac{2 \alpha}{1+\alpha  \mathcal X+\sqrt{1+2 \alpha   \mathcal X +\alpha^2 \mathcal Y^2}} 
\frac{  \cD_+ \Phi  \bar \cD_+ \bar \Phi    \p_\mm \bar \Phi  \p_\mm  \Phi}{  \sqrt{1+2 \alpha   \mathcal X +\alpha^2 \mathcal Y^2}}\,,
\\
    \cT_{----} &=-\frac{ \p_\mm\Phi   \p_\mm \bar\Phi }{ \sqrt{1+2 \alpha   \mathcal X +\alpha^2 \mathcal Y^2} }
    +(\cdots)\,\cD_+\Phi +(\cdots)\, \bar \cD_+\bar \Phi\,,
\end{aligned}
\end{equation}
where we indicated with ellipsis terms which will not play a role in our computation. Indeed when considering the product $
\cS_\pp \cT_{----}$ such terms vanish identically due to their Grassmann-odd nature and the fact that $\cS_{++}$
is proportional to $(\cD_+ \Phi  \bar\cD_+ \bar \Phi )$. 
Furthermore when truncating to the bosonic part of these supercurrents,  
one can check that the stress-energy tensor superfields  \eqref{descendants}  satisfy
\begin{equation}
\mathcal{T}_{\mu\nu}\Big|_{\text{bos.},\, \theta=0} = T_{\mu\nu}\,, 
\end{equation}
where $\mu,\nu=++,--$  and the right-hand side  is the bosonic stress-energy tensor in eq.~\eqref{eq:bosTmunu}.%
\footnote{Note that for the bosonic part of $\mathcal T_{----}$ to match $T_{----}$ we need the fermionic terms
 involving the ellipsis to vanish. We have verified that this is indeed the case on-shell, at leading order in~$\alpha$.}

From these expressions and eq.~\eqref{eq:W-R-relation} it follows that
\begin{equation}
\begin{aligned}
\bar \cW_-
&=\frac{-2\alpha}{1+\alpha  \mathcal X+\sqrt{1+2 \alpha   \mathcal X +\alpha^2 \mathcal Y^2}} 
\frac{ \p_\pp \bar \Phi  \cD_+ \Phi   \p_\mm \bar \Phi  \p_\mm  \Phi     }{ \sqrt{1+2 \alpha   \mathcal X +\alpha^2 \mathcal Y^2}}
+ (\cdots)\,\cD_+\Phi\, \cD_+\bar \Phi\,,~~~~~~
\\
   \cW_-&= \frac{-2\alpha}{1+\alpha  \mathcal X+\sqrt{1+2 \alpha   \mathcal X +\alpha^2 \mathcal Y^2}} 
   \frac{ \p_\pp  \Phi  \bar \cD_+ \bar \Phi   \p_\mm \bar \Phi  \p_\mm  \Phi  }{  \sqrt{1+2 \alpha   \mathcal X 
   +\alpha^2 \mathcal Y^2}} 
+ (\cdots)\,\cD_+\Phi\, \cD_+\bar \Phi\,,
\end{aligned}
\end{equation}
  where, once again, the terms in the ellipsis are irrelevant when considering the product $\bar \cW_-\,\cW_-$. 
It is now a matter of algebra to find
\bea
\mathcal O_{--}\,&=&   \cS_\pp \cT_{----}  - \bar \cW_- \cW_-
\non\\
&=&2 \frac{ \cD_+ \Phi  \bar\cD_+ \bar \Phi \p_\mm\Phi   \p_\mm \bar\Phi }
 {  \sqrt{1+2 \alpha   \mathcal X +\alpha^2 \mathcal Y^2} }
  \frac{1}{   1+\alpha   \mathcal X+\sqrt{1+2 \alpha   \mathcal X +\alpha^2 \mathcal Y^2} }\,,
\eea
which shows that indeed eq.~\eqref{eq:TTbarflowexplicit} holds.

\subsection{Expression in components and comparison with the ``Noether'' deformation}
It is instructive to rewrite the deformed action~\eqref{eq:susyTTbaraction} explicitly in components. 
By using the definition of the superfields~\eqref{eq:superfields} and performing some integration by parts we can recast 
the action in the form
\begin{equation}
\begin{aligned}
S_{\text{susy}} =\int \rd^2\sigma &\Bigg[A(x,y) + B(x,y)\Psi 
+ C(x,y)\, \partial_{--}\phi\partial_{--}\bar{\phi}\,\Psi_{++++}\\
&+D_{--}(x,y)\, \bar{\psi}_{+}\psi_{+} +E(x,y)\, \big(\Psi\big)^2 
+F(x,y)\, \partial_{--}\phi\partial_{--}\bar{\phi}\,\Psi\,\Psi_{++++}\\
& 
+G(x,y)\, \big(\partial_{--}\phi\partial_{--}\bar{\phi}\big)^2\,\big(\Psi_{++++}\big)^2
\Bigg]\,,
\label{susy-S-components}
\end{aligned}
\end{equation}
where the subscript ``susy'' emphasises that the action was obtained from our manifestly supersymmetric construction. 
Note that we introduced a short-hand notation for the fermion bilinears
\begin{equation}
\Psi = \bar{\psi}_+\partial_{--}{\psi}_+ + \psi_+\partial_{--}\bar{\psi}_+\,,\qquad
\Psi_{++++} = \bar{\psi}_+\partial_{++}{\psi}_+ + \psi_+\partial_{++}\bar{\psi}_+\,.
\end{equation}
The co\text{eff}icient $A(x,y), B(x,y)$, \textit{etc.}, depend on the bilinear combinations of the bosonic fields $x,y$ of 
eq.~\eqref{eq:xydef} and on the deformation parameter~$\alpha$; they are given in appendix~\ref{appComponent}. Without 
delving too deep in their specific form, we simply note that all these co\text{eff}icients are non-vanishing. We wish now to compare 
the form of this action with that of a $T\bar{T}$ deformation built out of the Noether energy-mometum tensor. As emphasised in 
ref.~\cite{Baggio:2018rpv}, we should not expect the two actions to be identical---indeed that was found \textit{not} to be the case 
already for deformations of an $\mathcal{N}=(0,1)$ Lagrangian. Again in ref.~\cite{Baggio:2018rpv}, the deformation of a free 
supersymmetric action of eight $\mathcal{N}=(1,1)$ multiplets was constructed by exploiting a connection with light-cone 
gauge-fixed strings~\cite{Baggio:2018rpv}. For the reader's convenience, let us copy that result---which is given in eq.~(4.18) 
there---specialising to the case where the undeformed action takes the form~\eqref{eq:freeaction02} corresponding to an 
$\mathcal{N}=(0,2)$ theory. We have
\begin{equation}
\begin{aligned}
S_{\text{Noether}}=&\int \rd^2\sigma\,
\frac{1}{2\alpha} \Bigg[
-1+2\ri\alpha\Psi 
\\
& +\sqrt{1+2\alpha x+\alpha^2 y^2
+\ri\alpha(4-\alpha x) \Psi  
-4\alpha^2 (\Psi )^2 -\ri\alpha^2 (\partial_{--}\phi\partial_{--}\bar{\phi})\Psi_{++++} }
\Bigg]\,.  \qquad\quad
\end{aligned}
\end{equation}
The action can be readily expanded in powers of the fermion bilinears $\Psi$ and $\Psi_{++++}$, and indeed it truncates at 
quadratic order. It is easy to see that no term without derivatives on the fermions---such as the one multiplying $D_{--}(x,y)$ in the 
supersymmetric action \eqref{susy-S-components}---may be generated in this expansion.%
\footnote{Indeed such a term cannot even be generated by partial integration, as that would introduce new fermion bilinears of the 
form~$\bar{\psi}_+\partial_{\pm\pm}{\psi}_+ - \psi_+\partial_{\pm\pm}\bar{\psi}_+$.}
Despite such a substantial difference, the two deforming operators, which may be found from the two actions by taking the
 partial $\alpha$-derivative, should coincide on-shell.%
 \footnote{Equivalently, the two theories should be the same up to (non-linear)
  field redefinitions. In particular, while the Noether-deformed action is not invariant under the free supersymmetry variations, 
  it should be invariant under suitably modified supersymmetry variations (whose form is induced by the non-linear field redefinition).}
  It is easy to verify that this is the case in the $\mathcal{N}=(0,1)$ case
  where $\psi_+\equiv\bar{\psi}_+$; then both actions are linear in $\Psi$ and $\Psi_{++++}$, and the $\bar{\psi}_+\psi_+$ 
  term vanishes identically. As discussed at some length in ref.~\cite{Baggio:2018rpv}, the fermion equations of motions for the 
  supersymmetric and ``Noether'' action then coincide, which is sufficient to show that the two deforming operators coincide 
  on-shell \textit{at all orders in}~$\alpha$. In the full $\mathcal{N}=(0,2)$ case, however, the fermion equations of motions are 
  different and in order to check the on-shell equivalence of the two deforming operators it is necessary to use both the fermion 
  and boson equations of motion which makes things rather less transparent. 
We have verified that the two deforming operators coincide on-shell up to order~$O(\alpha^3)$ and total derivatives
and we expect these results to hold at all order in $\alpha$.
Hence these two seemingly different deformations should give
rise to the same deformed theory, at least as long as the spectrum is
concerned. 
A more constructive check would be to explicitly produce the field redefinition relating the two actions. This would be particularly 
helpful in order to study the contact terms arising in deformed correlation functions. However, this  is a relatively difficult 
task. A simpler starting point would be the
$\mathcal{N} = (0, 1)$ case~\cite{Baggio:2018rpv,Chang:2018dge} 
before moving to the $\mathcal{N} = (0, 2)$ case.
We leave both these interesting studies, that fall beyond the scope of this paper, for future research.

 \section{Partial  supersymmetry breaking \texorpdfstring{$\cN=(2,2)\to(0,2)$}{N=(2,2)->(0,2)} }
 \label{sect-partial-breaking}
  
In the previous section, we have found the action for the supersymmetric $T\bar T$-deformation 
of a $\mathcal{N}=(0,2)$ free theory, eq.~\eqref{eq:susyTTbaraction}. 
This action, which is equivalent to the one originally studied in \cite{Hughes:1986dn}, as shown in \cite{Ivanov:2000nk},
resembles the four dimensional Bagger-Galperin action which describes the partial supersymmetry breaking from
 $\mathcal N=2$ to $\mathcal N=1$ in four dimensions~\cite{Bagger:1996wp}. 
Interestingly, the $T\bar T$-deformed free model is related to partial supersymmetry breaking 
in two dimensions.%
\footnote{Though it was not discussed in 
refs.~\cite{Baggio:2018rpv,Chang:2018dge}, it is simple to show that the
previously considered $T\bar{T}$ deformations of $\cN=(0,1)$ and $\cN=(1,1)$ free models of 
\cite{Baggio:2018rpv,Chang:2018dge} also possess additional, non-linearly realised supersymmetry ($\cN=(1,0)$ and
 $\cN=(1,1)$, respectively), as their actions are equivalent to those first studied in ref.~\cite{Ivanov:2000nk}.}
Indeed, following \cite{Ivanov:2000nk}, we will show that exactly the same action describes 
a model of partial supersymmetry 
breaking from $\mathcal N=(2,2) \rightarrow \mathcal N=(0,2)$ in two dimensions \cite{Ivanov:2000nk}.

In light-cone coordinates, 
a flat 2D $\cN=(2,2)$ superspace is parametrised by 
\begin{equation}
{\bm\z}^M=(\s^{\pm\pm},\q^\pm,\qb^\pm)\,,
\end{equation}
and spinor covariant derivatives and supercharges are given by
\bea
\cD_\pm=\frac{\p} {\p\theta^\pm}-\frac{\ri}{2} \bar \theta^\pm \p_{\pm\pm}
~, \qquad
\cQ_\pm=\frac{\p} {\p\theta^\pm}+\frac{\ri}{2} \bar \theta^\pm \p_{\pm\pm}
~, 
\eea
together with their complex conjugates.
They obey the anti-commutation relations
\bea
\{ \cD_\pm , \cDB_\pm \} =\ri \pa_{\pm\pm}\,,\qquad
\{ \cQ_\pm , \cQB_\pm \} =-\ri \pa_{\pm\pm}\,,
\eea
with all the other (anti-)commutators between $\cD$s, $\cQ$s, and $\pa_{\pm\pm}$ being identically zero.
Given an $\cN=(2,2)$ superfield  $\cF({\bm\z})=\cF(\s^{\pm\pm},\q^\pm,\qb^\pm)$ 
its supersymmetry transformations are given by
\bea
\d_Q \cF
:=
\ri\e^+ \cQ_+ \cF
+\ri\eta^- \cQ_- \cF
-\ri\bar\e^+ \cQB_+ \cF
-\ri\bar\eta^- \cQB_- \cF
~.
\label{susySuperfield22}
\eea

 To discuss the partial supersymmetry breaking, we begin introducing the simplest $\cN=(2,2)$ scalar multiplet.
 This is described by an $\cN=(2,2)$ chiral superfield $\Upsilon(\bm \z)$ satisfying 
 \be
\bar \cD_+ \Upsilon =\bar \cD_-\Upsilon =0
~.
 \ee
In general, $\Upsilon(\bm \z)$
 can efficiently be decomposed into $\cN=(0,2)$ multiplets expanding in the $\q^-$ and $\qb^-$ coordinates
\begin{equation}
\label{Upsilon}
\begin{aligned}
 \Upsilon(\bm \z)
 &=
 \Phi(\z)
+\theta^- \Psi^+(\z)
-\frac{\ri}{2} \theta^-\bar\theta^- \p_{--}\Phi(\z)\,,
\\
\Phi(\z)&:= \Upsilon(\bm \z)|_{\q^-=\qb^-=0}
\,,\\
\Psi^+(\z)&:=\cD_- \Upsilon(\bm \z)|_{\q^-=\qb^-=0}
\,.
\end{aligned}
\end{equation}
 Here $\Phi$ and  $\Psi^+$ are $\cN=(0,2)$ scalar and  Fermi chiral multiplets satisfying
  \be
\bar \cD_+\Phi =\bar \cD_+\Psi^+=0
\,.
 \ee

Since we are interested in partial supersymmetry breaking, 
we only consider the transformation under $\cQ_-,\bar \cQ_-$ and thus set $\epsilon^+=\bar \epsilon^+=0$. 
The $\epsilon^+$, $\bar \epsilon^+$ transformations will have preserved off-shell supersymmetry while
the $\eta^-$, $\bar \eta^-$ transformations will be the ones spontaneously broken. 
The supersymmetry transformation rules can straightforwardly be read from \eqref{susySuperfield22}
and, in particular, the two $\cN=(0,2)$ superfields transform under the left supersymmetry as
  \be\label{defsusytsf}
 \delta_\eta  \Phi=\ri\eta^- \Psi^+, \qquad \delta_\eta \Psi^+ =\bar \eta^- \p_{--} \Phi
 \,.
 \ee
To realize the partial supersymmetry breaking  
from $\cN=(2,2)$  to $\cN=(0,2)$, 
one needs to  deform the above transformation rules to
 \be
\tilde \delta_\eta  \Phi
 =
\ri\eta^- (\kappa \theta^++ \Psi^+)
\,, \qquad\tilde \delta_\eta \Psi^+ =\bar \eta^- \p_{--} \Phi
\,,
\label{susy20}
 \ee  
where  $\kappa$ has mass dimension 1 and represents the supersymmetry breaking scale. 
As described in details in \cite{Hughes:1986dn,Ivanov:2000nk},
the extra $\kappa\eta^-  \theta^+$ term is linked to a central charge deformation of the supersymmetry algebra
which is necessary to have partial supersymmetry breaking. 
In fact, if we define $\Xi_+:=\cD_+ \Phi$, 
by using the previous transformations it follows
 \be
\tilde\delta_\eta \Xi_+ =-\ri\eta^- (\kappa + \cD_+\Psi^+) 
\,.
 \ee
As in \cite{Wess:1992cp},
these transformations can be converted into  standard non-linearly realised supersymmetry transformations  by defining 
 \be
 \tilde \Xi_+ =e^{\tilde\delta_\eta} \Xi_+ \Big|_{\eta=-\frac1\kappa \lambda}~, \qquad
   \tilde  \Psi^+ =e^{\tilde\delta_\eta}  \Psi^+\Big|_{\eta=-\frac1\kappa \lambda}~,
 \ee
 where $\lambda$, which transforms as
 \be
\tilde \delta_\eta \lambda^-= \kappa \eta^-  - \frac{\ri}{2\kappa} ( \eta^- \bar \lambda^- +\bar \eta^- \lambda^-)\p_{--} \lambda^-
 \,,
 \ee
 is the complex goldstino associated to the $\cN=(2,2)\to(0,2)$ partial supersymmetry breaking.
One can then show that $\tilde\Xi_+$ and $\tilde \Psi^+$ transform homogeneously  as
  \be
 \tilde \delta_\eta \tilde\Xi^+ =  - \frac{\ri}{2\kappa} ( \eta^- \bar \lambda^- +\bar \eta^- \lambda^-)\p_{--}  \tilde\Xi^+, \qquad
\tilde  \delta_\eta \tilde \Psi^+ =  - \frac{\ri}{2\kappa} ( \eta^- \bar \lambda^- +\bar \eta^- \lambda^-)\p_{--} \tilde \Psi^+ 
  \,.
 \ee
This enables one to impose the supersymmetric invariant constraints  $\tilde\Xi_+=\tilde \Psi^+ =0$
that are solved by
\be\label{Psieq}
 \Psi^+=\ri \frac{ \bar\cD_+\bar \Phi \p_{--} \Phi  }{\kappa -  \bar\cD_+ \bar \Psi^+}
 \,, \qquad
  \bar \Psi^+=-\ri \frac{   \cD_+ \Phi \p_{--}\bar \Phi  }{\kappa + \cD_+ \Psi^+}
 \,, \qquad
 (\Psi^+)^2=(\bar\Psi^+)^2=0
 \,.
 \ee
 The previous result can  also be rewritten as 
\begin{equation}
\label{PsiDpsi}
 \Psi^+ =\frac{1}{\kappa}\bar \cD_+ \Big[ \ri\bar\Phi \p_{--} \Phi -\Psi^+ \bar\Psi^+\Big]
 =\frac{1}{\kappa}\bar \cD_+ \Bigg[ \ri\bar\Phi \p_{--} \Phi 
 -   \frac{ \bar\cD_+\bar \Phi   \cD_+ \Phi \p_{--} \Phi\p_{--}\bar \Phi  }{(\kappa -  \bar\cD_+ \bar \Psi^+)(\kappa + \cD_+ \Psi^+)} 
 \Bigg]
 \,,
\end{equation}
 together with its complex conjugates.
 By using a standard trick \cite{Bagger:1996wp,Ivanov:2000nk}, 
 the denominator  $(\kappa + \cD_+ \Psi^+)$ cannot contribute terms like 
 $\bar \cD_+\bar \Phi$, since the same fermionic terms appear already in the numerator. Hence the 
 $(\kappa + \cD_+ \Psi^+)$  term only appears \text{eff}ectively as
 \be
 (\kappa + \cD_+ \Psi^+)_{\text{eff}}=
 \Big (\kappa + \cD_+  \frac{\ri \bar\cD_+\bar \Phi \p_{--} \Phi  }{\kappa -  \bar\cD_+ \bar \Psi^+}  \Big)_{\text{\text{eff}}} 
= \kappa -     \frac{\p_{++} \bar\Phi \p_{--} \Phi  }{\kappa - ( \bar\cD_+ \bar \Psi^+)_{\text{eff}}}    
\,,
\ee
which leads to
\be
( \cD_+ \Psi^+)_{\text{eff}} = -     \frac{\p_{++} \bar\Phi \p_{--} \Phi  }{\kappa - ( \bar\cD_+ \bar \Psi^+)_{\text{eff}}}   
\,,
\qquad
( \bar \cD_+\bar \Psi^+)_{\text{eff}} =       \frac{\p_{++}  \Phi  \p_{--}  \bar\Phi  }{\kappa +(  \cD_+   \Psi^+)_{\text{eff}}}   
\,.
\ee
Their solution gives
\begin{equation}
\begin{aligned}
( \cD_+ \Psi^+)_{\text{eff}} &=
\frac{1}{2\kappa} \Big(B-\bar B-\kappa^2 +\sqrt{\kappa^4 - 2\kappa^2(B+\bar B) +(B-\bar B)^2}  \Big)
\,,
\\
B &:=\pa_{++}\Phi\pa_{--}\bar{\Phi}
\,,\\
\bar{B} &:= \pa_{++}\bar\Phi\pa_{--}\Phi
\,.
\end{aligned}
\end{equation}
Substituting back into \eqref{Psieq}, one can express $\Psi^+$ in term of $\Phi,\, \bar \Phi$ and their derivatives
 \be 
 \Psi^+  =\frac{1}{\kappa}\bar \cD_+ \Bigg[ \ri\bar\Phi \p_{--} \Phi 
 -   \frac{2 \bar\cD_+\bar \Phi   \cD_+ \Phi \p_{--} \Phi      \p_{--}\bar \Phi  }
 { \kappa^2-\cX+    \sqrt{\kappa^4 - 2\kappa^2\cX +\cY^2}  } \Bigg]
 \,.
 \label{PsiDpsi-2}
 \ee

 By construction, the two  $\cN=(0,2)$  superfields $\Phi(\z)$ and $\Psi^+(\z)$ also
 possess a hidden non-linearly realised $\cN=(2,0)$ supersymmetry \eqref{susy20}.  
 
Now we can construct the following full superspace action 
\bsubeq
 \beqn
S_{\kappa}&= &
 \frac{1}{2}\int \rd^2\s\,\rd\theta^+ \rd\bar\theta^+  \rd\theta^- \rd\bar\theta^- \Upsilon \bar\Upsilon 
  \label{super1}
\\
 & =&
  \frac{1}{2}
 \int \rd^2\s\,\rd\theta^+ \rd\bar\theta^+ \Big[  \frac{\ri}{2}(\bar \Phi \p_{--} \Phi-\Phi \p_{--}\bar\Phi)
   -  \Psi^+   \bar\Psi^+ \Big]
   \label{sk1}
 \,.
\eeqn
\esubeq
Alternatively, since $\Upsilon$ is chiral, one can also consider the following supersymmetric action integrating over half 
superspace
\bsubeq
 \beqn
S_{\kappa} &=&
 -\frac{\kappa}{4} \int \rd^2\s\,\rd \theta^+\rd\theta^- \Upsilon 
 +\frac{\kappa}{4} \int \rd^2\s\,\rd \bar \theta^+\rd \bar \theta^- \bar \Upsilon
 \label{super2}
\\&=&
-\frac{\kappa}{4} \int \rd^2\s\,\rd \theta^+\Psi^+
-\frac{\kappa}{4} \int \rd^2\s\,\rd \bar \theta^+\bar\Psi^+
\label{sk2}
  ~.
\eeqn
\esubeq
Note that here $\Upsilon(\bm\z)$ is defined as in \eqref{Upsilon} but with $\Phi(\z)$ and $\Psi^+(\z)$ 
now transforming as in \eqref{susy20}.
Then the supersymmetry $\cN=(2,2)$ transformations of $\Upsilon(\bm\z)$ gets modified to
\bea
\tilde\d_Q\Upsilon
=
\d_Q\Upsilon
+\ri\kappa \eta^-  \theta^+
~,
\eea
with $\d_Q\Upsilon$ as in \eqref{susySuperfield22}.
Despite the deformation of   the $\mathcal N=(2,0)$   supersymmetry one can still explicitly verify that   the $\mathcal N=(2,0)$ 
supersymmetric variations of the integrands in \eqref{sk1} and \eqref{sk2} are total derivatives  \cite{Ivanov:2000nk}. 
 Together with their manifest $\mathcal N=(0,2)$ supersymmetry,  one finds that  \eqref{sk1} and \eqref{sk2}  are supersymmetric 
 under $\mathcal N=(2,2)$. This also justifies the manifest $\mathcal N=(2,2)$  superspace formulation of the actions in 
 \eqref{super1} and \eqref{super2}.

Using \eqref{PsiDpsi}, it can be even shown that the above two actions  with   $\mathcal N=(2,2)$ 
supersymmetry are equivalent  
\begin{equation}
\begin{aligned}
S_{\kappa}&=
 \frac{1}{2}\int \rd^2\s\,\rd\theta^+ \rd\bar\theta^+  \rd\theta^- \rd\bar\theta^- \Upsilon \bar\Upsilon 
\\
&=-\frac{\kappa}{4} \int \rd^2\s\,  \rd \theta^+\rd\theta^-
 \Upsilon +\frac{\kappa}{4} \int \rd^2\s\, \rd \bar \theta^+\rd \bar \theta^- \bar \Upsilon  \,.
\end{aligned}
\end{equation}
Inserting the  explicit solution  \eqref{PsiDpsi-2}, one can explicitly write down the action as 
\begin{equation}
\begin{aligned}
S_{\kappa}  &=-\frac{\kappa}{2} \int \rd^2\s\, \rd \theta^+\Psi^+   
 = \int \rd^2\s\,\rd\theta^+ \rd\bar\theta^+  \Bigg[ \frac{\ri}{2}\bar\Phi \p_{--} \Phi 
+ \frac{ \cD_+ \Phi \bar\cD_+\bar \Phi    \p_{--} \Phi      \p_{--}\bar \Phi  }
 { \kappa^2- \mathcal X+    \sqrt{ \kappa^4 -2 \kappa^2 \mathcal X  + \mathcal Y^2\big.}  } \Bigg]
 \,.
 \label{Skappa}
\end{aligned}
\end{equation}
Once we identify 
 \be
 \alpha =-\frac{1}{\kappa^2}
 \,,
 \ee
it is obvious that  \eqref{Skappa} gives exactly our previous 
$T\bar T$-deformed action $S_\alpha$ in eq.~\eqref{eq:susyTTbaraction}, 
showing that $S_\alpha$, besides being manifestly $\cN=(0,2)$ supersymmetric,
is also invariant under the extra spontaneously broken $\cN=(2,0)$ supersymmetry \eqref{susy20}.

\section*{Acknowledgements}
We thank Marco Baggio for discussions and collaboration at the initial stages of this project.
We also thank Chih-Kai Chang, Christian Ferko and Savdeep Sethi for discussions and collaboration on related projects
and for comments on the manuscript.
AS and GT-M thank the participants of the workshop 
\emph{$T\bar{T}$ and Other Solvable Deformations of Quantum Field Theories} 
for the stimulating atmosphere  and the Simons Center for Geometry and Physics for hospitality and partial support 
during the final stages of this work.
 This work is partially supported through a research grant of the Swiss National Science Foundation, 
 as well as by the NCCR SwissMAP, funded by the Swiss National Science Foundation.
 AS's work was supported by ETH Career Seed Grant SEED-23 19-1.
The work of GT-M is supported by the Albert Einstein Center for Fundamental Physics, University of Bern,
and by the Australian Research Council (ARC) Future Fellowship FT180100353.
  
\newpage
 \appendix

 \section{Useful results for section \ref{point-plitting-section}}
\label{appendix-for-point-splitting-section}
Here we collect some useful technical results used in section \ref{point-plitting-section}.
Equation \eqref{pammO-0_pappO-00} are
\bsubeq
\bea
\pa_{--}\cO_{--}(\z,\z')
&=&
-\cT_{----}(\z) \left[\pa'_\mm\cS_\pp(\z') - \cD'_+ \cW_-(\z') +\bar \cD'_+ \bar \cW_-(\z')\right]
\non\\
&&
+\left[\cD_+\cT_{----}(\z)-\hf\pa_\mm\bar\cW_-(\z)\right] \cW_-(\z')
\non\\
&&
-\left[\cDB_+\cT_{----}(\z)-\hf \pa_\mm\cW_-(\z)\right] \bar \cW_-(\z')
\non\\
&&
+(\pa_{--}+\pa'_{--})\left[\cT_{----}(\z)\cS_\pp(\z')   \right]
\non\\
&&
-(\cD_++\cD'_+)\left[\cT_{----}(\z) \cW_-(\z')\right]
\non\\
&&
+(\cDB_++\cDB'_+)\left[\cT_{----}(\z) \bar \cW_-(\z')\right]
~,
\label{pammO}
\eea
and
\bea
\pa_{++}\cO_{--}(\z,\z')
&=& 
-\ri \Big[\cD_+ \Big(\cDB_+\cT_{----}(\z)- \frac12 \pa_{--}  \cW_-(\z)\Big)\Big]\cS_\pp(\z')
\non\\
&&
-\ri \Big[\cDB_+\Big( \cD_+\cT_{----}(\z) -\frac12 \pa_{--} \bar\cW_- (\z)\Big)\Big]\cS_\pp(\z')
\non\\
&&
+\frac{\ri}{2}\left[ \cD_+\cW_{-}(\z)+ \cDB_+\bar{\cW}_-(\z)\right]
\left[\pa'_\mm\cS_\pp(\z') - \cD'_+ \cW_-(\z') +\bar \cD'_+ \bar \cW_-(\z')\right]
\non\\
&&
-\frac{\ri}{2}(\pa_{--}+\pa'_{--})\left[ \cD_+\cW_{-}(\z)\cS_\pp(\z')+ \cDB_+\bar{\cW}_-(\z)\cS_\pp(\z')\right]
\non\\
&&
-\frac{\ri}{2}( \cDB_++\bar \cD'_+)\left[\big( \cD_+\cW_{-}(\z)\big) \bar \cW_-(\z')\right]
\non\\
&&
+\frac{\ri}{2}( \cD_++ \cD'_+) \left[\big(\cDB_+\bar{\cW}_-(\z)\big) \cW_-(\z') \right]
\non\\
&&
+\frac{\ri}{2}( \cD_++ \cD'_+) \left[\cW_{-}(\z) \cD'_+\cW_-(\z') \right]
\non\\
&&
-\frac{\ri}{2}( \cDB_++ \cDB'_+) \left[\bar{\cW}_-(\z)\bar \cD'_+\bar \cW_-(\z')\right]
~,~~~~~~~~~
\label{pappO}
\eea
\esubeq
where the unprimed superspace covariant derivatives, $\cD_A=(\pa_{\pm\pm},\cD_+,\cDB_+)$, 
act only on the superspace coordinates $\z$ while the primed derivatives $\cD'_A$ act only on $\z'$.
To emphasise the difference between \eqref{pammO-0_pappO-00}, \eqref{pammO}, \eqref{pappO} 
and \eqref{pammO-0_pappO-0}, before and after the 
$\q=\q'$ limit,
it is also useful to point out that the following equations hold:
\bsubeq\label{DQ}
\bea
( \cD_++\cD'_+) &=& 
(\cQ_++\cQ'_+) 
-\ri\qb^+ (\pa_{++}+\pa'_{++})
+\ri(\qb^+-\qb'{}^+) \pa'_{++}
~,
\\
( \cD_++\cD'_+) &=& 
(\cQ_++\cQ'_+) 
-\ri\qb'{}^+ (\pa_{++}+\pa'_{++})
-\ri(\qb^+-\qb'{}^+) \pa_{++}
~,
\eea
\esubeq
and
\bsubeq\label{DQb}
\bea
( \bar \cD_+ +\bar \cD'_+) 
 &=& 
(\bar\cQ_++\bar\cQ'_+)
 + \ri\q^+ (\pa_{++}+\pa'_{++})
- \ri(\q^+-\q'{}^+) \pa'_{++}
~,
\\
( \bar \cD_+ +\bar \cD'_+) 
 &=& 
(\bar\cQ_++\bar\cQ'_+)
 + \ri\q'{}^+ (\pa_{++}+\pa'_{++})
+ \ri(\q^+-\q'{}^+) \pa_{++}
~.
\eea
\esubeq
The last terms, that are function of the distances in the grassmannian directions, $(\q^+-\q'{}^+)$ and $(\qb^+-\qb'{}^+)$,
are not multiplying a generator of (super-)translations. For this reasons, in general, they do not annihilate superspace OPE
coefficients, which is necessary for the argument in section \ref{point-plitting-section} to go through.
Since these terms disappear when $\q=\q'$, 
it is enough to consider the coincident grassmannian limit for \eqref{pammO-0_pappO-0}, 
which derives from \eqref{pammO}--\eqref{DQb}, to be true.
This is in the end suffices to show the well-definedness of the composite operator $\cO_{--}(\z)$.
To make more clear how to properly read eq.~\eqref{pammO-0_pappO-0} let us elaborate further on how to interpret 
expressions where the $\q=\q'$ limit is taken.

Given two superfields $\cU^1(\z)$ and $\cU^2(\z')$ we consider the superspace point-split
bilocal operator $\cO(\z,\z')=\cO(\s,\q;\s',\q')$ defined as
\bea
\cO(\z,\z')
=
\cU^1(\z)\cU^2(\z')
~.
\eea
Its $\q=\q'$ limit, $\cO(\s,\s',\q):=\cO(\s,\q;\s',\q)$, is 
\bea
\cO(\s,\s',\q)
=
\cU^1(\s,\q)\cU^2(\s',\q)
~,
\eea
and represents a point-split version in the bosonic coordinates $\s$ and $\s'$ of the composite operator
$\cO(\z):=\cU^1(\z)\cU^2(\z)$.
We define  the following differential operators
\bsubeq
\bea
\widehat{\cD}_+&:=&\frac{\p} {\p\theta^+}-\frac{\ri}{2} \bar \theta^+ \widehat{\p}_{++}~, \qquad
\overline{\widehat{\cD}}_+=-\frac{\p} {\p\bar \theta^+}+\frac{\ri}{2}   \theta^+ \widehat{\p}_{++}
~,
\\
\widehat{\cQ}_+&:=&\frac{\p} {\p\theta^+}+\frac{\ri}{2} \bar \theta^+ \widehat{\p}_{++}~, \qquad
\overline{\widehat{\cQ}}_+=-\frac{\p} {\p\bar \theta^+}-\frac{\ri}{2}   \theta^+ \widehat{\p}_{++}
~,
\eea
\esubeq
with
\bea
\widehat{\p}_{\pm\pm}&:=&\frac{\pa}{\pa \s^{\pm\pm}}+\frac{\pa}{\pa\s'{}^{\pm\pm}}
~.
\eea
These satisfy the same algebra as the un-hatted covariant derivatives and supercharges, e.g.
$\{\widehat{\cD}_+,\overline{\widehat{\cD}}_+\}=\ri\widehat{\pa}_{++}$ etc.
It is then clear that, thanks to the Leibniz rule of the spinor derivatives $\frac{\pa}{\pa\q^+}$ and $\frac{\pa}{\pa\qb^+}$, it holds
\bsubeq
\bea
\widehat{\cD}_+\cO(\s,\s',\q)&=&\Big\{ (\cD_++\cD'_+)\cO(\z,\z')\Big\}\big|_{\q=\q'}
~,
\\
\widehat{\cQ}_+\cO(\s,\s',\q)&=&\Big\{ (\cQ_++\cQ'_+)\cO(\z,\z')\Big\}\big|_{\q=\q'}
~,
\eea
\esubeq
and similar expressions for their complex conjugates.
The convenience to have introduced the hatted operators becomes clear when we consider
how supersymmetry transformations act on $\cO(\s,\s',\q)$.
By using \eqref{susySuperfield} for  $\d_Q\cU^1(\z)$ and $\d_Q\cU^2(\z)$,
it follows
\bsubeq
\bea
\d_Q\cO(\s,\s',\q)
&=&
[\d_Q\cU^1(\s,\q)]\cU^2(\s',\q)
+\cU^1(\s,\q)[\d_Q\cU^1(\s',\q)]
\non\\
&=&
\Big\{\big[
\ri\e^+\big( \cQ_++\cQ'_+\big)
-\ri\bar{\e}^+\big( \cQB_++\cQB'_+\big)
\big]\cO(\z,\z')
\Big\}\big|_{\q=\q'}
\\
&=&
\Big\{\big[
\ri\e^+\big( \cD_++\cD'_+\big)
-\ri\bar{\e}^+\big( \cDB_++\cDB'_+\big)
\big]\cO(\z,\z')
\Big\}\big|_{\q=\q'}
\non\\
&&
+(\e^+\qb^+ +\bar{\e}^+\q^+) (\pa_{++}+\pa'_{++})\cO(\s,\s',\q)
~,
\eea
\esubeq
which can be equivalently represented as
\bsubeq
\bea
\d_Q\cO(\s,\s',\q)
&=&
\big[\ri\e^+\widehat{\cQ}_+-\ri\bar{\e}^+\overline{\widehat{\cQ}}_+\big]\cO(\s,\s',\q)
\\
&=&
\big[\ri\e^+\widehat{\cD}_+-\ri\bar{\e}^+\overline{\widehat{\cD}}_+\big]\cO(\s,\s',\q)
+(\e^+\qb^+ +\bar{\e}^+\q^+) \widehat{\pa}_{++}\cO(\s,\s',\q)
~.
~~~~~~
\eea
\esubeq
Then the hatted operators are the ones generating translations and $\cN=(0,2)$ supersymmetry transformations
of bilocal operators such as $\cO(\s,\s',\q)$. These are the operators appearing in \eqref{pammO-0_pappO-0}.
Moreover, the results above  make it evident that  taking the $\q=\q'$ limit and acting on 
bilocal superfields with Grassmann dependent differential operators are two  commuting operations.


 \section{Computation of the supercurrent}
 \label{sugra-supercurrent}
 
In this appendix,  we calculate the supercurrent of the action  \eqref{eq:susyTTbaraction}   
in order to verify that it is indeed arising form the $T\bar T$ deformation of the free theory \eqref{eq:freeaction02}. 
The strategy to compute the supercurrent is to couple the model to supergravity in superspace
and then taking functional derivates with respect to the gravitational superfield prepotentials.
This procedure is the superspace analogue of the calculation of the  Hilbert stress-energy tensor in a generic QFT. 

The study of 2D $\cN=(0,2)$ supergravity in superspace was largely developed in the 1980s
and we refer the reader to the following works and references therein for details 
\cite{Brooks:1986gd,Brooks:1986uh,Brooks:1987nt,Evans:1986ada,Evans:1986tc,Evans:1986wp,Evans:1986wq,Govindarajan:1991sx}. 
In particular,
we refer  to  \cite{Brooks:1987nt,Govindarajan:1991sx} that we will closely follow including their notations. 
The covariant derivatives are defined as
 \be\label{sugranotation}
\cD_+=\frac{\p}{\p \theta^+} +\ri\bar \theta^+ \p_{++}
\,,\qquad  \bar\cD_+=\frac{\p}{\p \bar \theta^+} +\ri  \theta^+ \p_{++}
~,
\ee
satisfying 
\be
\cD_+^2=\bar \cD_+^2=0\,, \qquad  \{\cD_+ , \bar \cD_+ \}=2\ri \p_{++}\,,
 \qquad   \{\cD_+ , \p_{\pm \pm} \}=\{\bar\cD_+ , \p_{\pm \pm} \}=0
 ~.
\ee

Due to the covariant properties of these derivatives and the isomorphism of different representations of the superalgebra, 
all the expressions in different notations should take the same form except for the coefficients. When translating among different 
notations, the coefficients can be fixed unambiguously by comparing the component expression. For our superrcurrent, the 
coefficients can be fixed by considering the component of the suprcurrent which are related to the energy-momentum tensor.

 \subsection{\texorpdfstring{$\cN=(0,2)$}{N=(0,2)} supergravity}
 
 In this section, we review the 2D $\mathcal N=(0,2) $ supergravity following references 
 \cite{Brooks:1987nt,Govindarajan:1991sx}.  
The superspace geometry we consider is based on a structure group based on the 2D Lorentz group.
The covariant derivatives include the super-Vielbein $E_M{}^A$ and its inverse $E_A{}^M$, together with
the Lorentz connection superfield $\omega_M$, 
and take in general the form%
\footnote{The $\bar{+}$ notation is used for convenience only
 to keep track of barred and unbarred terms.} 
$\nabla_A=(\nabla_{\pm\pm},\nabla_+,\bar\nabla_{\bar{+}})$
\be
\nabla_A=E_A{}^M \pa_M+\omega_A \mathcal M
~,
~~~~~~
\pa_M=\Big(\frac{\pa}{\pa\s^{\pm\pm}},\frac{\pa}{\pa\q^{+}},\frac{\pa}{\pa\qb^{\bar{+}}}
\Big)
~.
\ee 
They satisfy an algebra of the form
\bea
[\nabla_A,\nabla_B\}
=
T_{AB}{}^C\nabla_C
+R_{AB}\cM
~.
\eea
The torsion $T_{AB}{}^C$ 
and curvature $R_{AB}$
superfields represent highly reducible representations of local supersymmetry and they are in general 
constrained to appropriately describe the multiplet of $\cN=(0,2)$ Poincar\'e supergravity off-shell.
We refer to \cite{Brooks:1987nt,Govindarajan:1991sx} for a detailed analysis of the constraints for the torsion and
curvature tensors and the Bianchi identities they satisfy. 
For the purpose of computing the supercurrent 
it is enough to describe how the constraints are solved at the linear order in terms of 
 a set of unconstrained ``prepotential'' superfields that play the role of the metric in the context of superfield supergravity,
 see \cite{Gates:1983nr,Buchbinder:1998qv} 
 for pedagogical reviews. 
At linearised order the covariant derivatives  can be expanded about a flat background as
 \be
\nabla_A=\cD_A-H_A{}^M  \cD_M+\omega_A(H) \mathcal M
~,
\ee
where the superconnection $\o_A$ is completely determined in terms of $H_A{}^M$.
To linear order, the constraints can be solved in terms of three independent prepotential superfields: 
$H_{--}{}^{--}$, $H_{--}{}^{++}$, and $H^{--}$.  
All the other components of  $H_A{}^M$ can be expressed in terms of the  prepotentials. The expressions used in our paper are%
\footnote{The complex conjugate relation is $(H_A{}^M )^*= H_{\bar A }{}^{\bar M}  (-)^{|A|+|M|}$.}
\bsubeq
   \beqn
 H_{--}{}^{ +  } &= & \frac{1}{2\ri} \bar \cD_+  H_{--}{}^{++} 
 ~,   \\
 H_{--}{}^{\bar +  } &= &- \frac{1}{2\ri}  \cD_+  H_{--}{}^{++}    
 ~,
\\
 H_{++}{}^{+  } &= & \frac{1}{2\ri} \bar \cD_+  H_{--}{}^{--}   
 ~, \\
  H_{++}{}^{ \bar +  } &= &- \frac{1}{2\ri}   \cD_+  H_{--}{}^{--}   
  ~, \\
 H_{++}{}^{+ +} &= &  H_{--}{}^{--}  
 ~, \\
 H_{++}{}^{--} &= &  -\frac12 [\cD_+, \bar \cD_+]H ^{--}    
 ~,
\\
 H_+{}^{\bar +} &= &  H_+{}^{++} =  H_{\bar+}{}^{  +} =  H_{\bar+}{}^{++} = 0  
 ~,\\
   H_+{}^{ +} +   H_{\bar +}{}^{\bar +} &= & H_{--}{}^{--}   
   ~,\\
 H_+{}^{--} &= &  \ri \cD_+ H^{--}   
 ~,\\
 H_{\bar+}{}^{--} &= & - \ri \bar\cD_+ H^{--}   
 ~.
 \eeqn
\esubeq

The various components of the supergravity multiplet can be obtained from the prepotentials through projections. 
In particular, the linearised metric fluctuations are
\bsubeq
\bea
h_{--}{}^{++}&=&- H_{--}{}^{++}|_{\theta=0}~,
\\
h_{--}{}^{--}&=&h_{++}{}^{++}=- H_{--}{}^{--}|_{\theta=0}
~, 
\\
h_{++}{}^{--}&=& \frac12 [\cD_+, \bar \cD_+] H^{--}|_{\theta=0}
~.
\eea
\esubeq

 After solving the constraints, the linearised supergravity transformations of the prepotentials turn out to be
  \cite{Brooks:1987nt,Govindarajan:1991sx}
\bsubeq
 \beqn
 \delta H^{--}& = &\ri(  \Lambda_{++}-  \bar \Lambda_{++} )
 ~,
  \\
 \delta H_{--}{}^{--}  & = &- \frac12 \p_{--}( \Lambda_{++} +  \bar \Lambda_{++}  )   -\frac12 \p_{++} K_{--} 
 ~,
  \\
 \delta H_{--}{}^{++}  & = &- \p_{--} K_{--}
 ~,
 \eeqn
\esubeq
 where $K_{--}$ is real,  while $  \Lambda_{++}, \bar   \Lambda_{++} $ are chiral and anti-chiral respectively:
 \be
 K_{--}=\bar K_{--}
 ~, \qquad 
 \bar  \cD_+   \Lambda_{++}=0
 ~, \qquad 
 \cD_+\bar   \Lambda_{++}=0
 ~.
 \label{sugra-transf}
 \ee

\subsection{\texorpdfstring{$\cR$}{R}-multiplet from  supergravity}
 
Consider a general Lorentz invariant matter system coupled to $\mathcal{N}=(0,2)$ supergravity. 
Its action expanded to first order in the supergravity prepotential is
 \be
 S_{ \rm int}=-\frac14 \int \rd^2\sigma \,\rd\theta^+ \rd\bar \theta^+\;  
 \Big( 2H^{--} \cT_{----} +2H_{--}{}^{--} \cR_{--}+ H_{--}{}^{++}\cR_{++}  \Big)
 ~,
 \ee
which leads to
 \be
  S_{ \rm int}=-\frac12 \int \rd^2 \sigma  \Big(  h_{--}{}^{++} T_ {++++}+2h_{--}{}^{--}T_{ ++--}+h_{++}{}^{--}T_{----}   \Big) +
  \cdots~
 \ee
 where the ellipsis represent the fermionic contributions.
 
Assuming that the equations of motion for the matter are satisfied,
the variation of the action under arbitrary supergravity gauge transformations \eqref{sugra-transf} takes the form
 \beqn
 \delta S_{\rm int} = - \frac14
  \int \rd^2\sigma \,\rd\theta^+ \rd\bar \theta^+&\Big\{&
2 \ri  \Lambda_{++} \Big(\cT_{----}  - \frac{\ri}{2} \p_{--} \cR_{--} \Big)
  - 2\ri \bar \Lambda_{++}\Big(\cT_{----}  +\frac{\ri}{2} \p_{--} \cR_{--} \Big)
 \non \\
 &&
  +   K_{--}\big(\p_{++} \cR_{--}+\p_{--}\cR_{++}\big)
 \Big\}
 ~.
 \eeqn
 The invariance of the action then dictates the following conservation equations
\bsubeq
 \beqn
 \bar\cD_+\Big( \cT_{----}  -\frac{\ri}{2} \p_{--} \cR_{--} \Big) &=& 0 
 ~, \\
 \cD_+\Big( \cT_{----}  + \frac{\ri}{2} \p_{--} \cR_{--} \Big) &=& 0  
 ~,\\
 \p_{++} \cR_{--}+\p_{--}\cR_{++} &=&0 
 ~.
 \eeqn
\esubeq
 These, are exactly the conservation law for the $\cR$-multiplet \eqref{Rmultiplet}.

 \subsection{Computing the supercurrent}
 
Next we are going to derive the supercurrent multiplet for the models of interest in our paper. 
To do that, we first need to covariantise the actions. 

Since we are dealing with scalar multiplets, 
 the Lorentz connection $\omega_A$ will be irrelevant for our calculations. 
 Then the  covariant derivatives will always be
\be
\nabla_A= \cD_A-H_A{}^M  \cD_M
~.
\ee 
The superdensity is expanded at the linear order in  terms of the prepotentials as
\be
E^{-1}=1+\text{Str} H_A{}^M=
1+H_{++}{}^{++}+H_{-- }{}^{--}-H_{+}{}^{+}-H_{-}{}^{-}=1+H_{-- }{}^{--}
~.
\ee
We also need to define the covariantly chiral and anti-chiral superfields in supergravity 
\be
\nabla_+  \bar {\hat \Phi}=(\cD_+-H_{+}{}^{+} \cD_+-\ri \cD_+ H^{--} \p_{--}   ) \bar {\hat \Phi}=0
~.
\ee 
To linearised order, one finds the following expression
for a covariantly (anti-)chiral superfield, ($\bar{\hat{\Phi}}$) $\hat{\Phi}$, in terms of a standard
(anti-)chiral superfield, ($\bar{\Phi}$) $\Phi$, 
  \be
 {\hat \Phi}= (1-\ri H^{--} \p_{--} )\ \Phi
 ~,~~~~~~
 \bar {\hat \Phi}= (1+\ri H^{--} \p_{--} )\bar\Phi
 ~.
 \ee
To covariantise a matter action coupled to supergravity, 
we replace all the quantities with covariant ones: 
$\cD_A \rightarrow \nabla_A$, 
$\Phi\rightarrow\hat \Phi, \bar \Phi \rightarrow  \bar {\hat \Phi} $. 
The superdensity $E^{-1}$ should also be taken into account in the superspace measure.

\subsubsection{Free theory}

To illustrate the strategy of computing the supercurrent,  
let us first consider the free theory. The action is 
  \be
 S_0= -\frac{\ri}{4}  \int \rd^2\sigma\, \rd\theta^+ \rd\bar\theta^+  \,\bar\Phi \p_{--} \Phi  \,
 ~.
 \ee
In the supergravity case the action takes the form
 \be
 S_0=-\frac{\ri}{4}   \int \rd^2\sigma\, \rd\theta^+ \rd\bar\theta^+ E^{-1}  \bar{\hat\Phi} \nabla_{--}  \hat\Phi   
 ~.
 \label{curved-free}
 \ee
The integrand can be computed explicitly.  To linear order in the prepotentials, it holds
 \beqn  
 E^{-1}  \bar{\hat\Phi} \nabla_{--}  \hat\Phi 
  &=&
    \bar\Phi \p_{--} \Phi  +H_{-- }{}^{--}  \big( \bar\Phi\p_{--}\Phi -  \bar\Phi\p_{--}\Phi  \big)
    \non\\
    &&
    + \big( -H_{-- }{}^{++}  \bar \Phi \p_{++}\Phi +\frac{\ri}{2} \bar \cD_+ H_{--}{}^{++} \bar\Phi \cD_+ \Phi  \big)
   \qquad   
         \non\\& &
      + \big(  \, \ri H^{--} (\p_{--} \bar\Phi \p_{--}\Phi  -\bar\Phi \p_{--}^2\Phi) -\ri \bar\Phi \p_{--} H^{--} \p_{--}\Phi  \big)
~. \eeqn
 Plugging this  back into the action \eqref{curved-free} 
 and integrating by parts, we get 
  \be
 S_0
 =
 -\frac{1}{4} \int \rd^2\sigma\, \rd\theta^+ \rd\bar\theta^+  \Big( 
\ri\bar\Phi \p_{--} \Phi 
+ 2H^{--} \cT_{----} 
+2H_{--}{}^{--} \cR_{--}
+H_{--}{}^{++}\cR_{++}
    \Big)
    ~.
 \ee
 where 
\bsubeq
 \beqn
\cR_{-- } &=& 0 
~,\\
\cR_{++ }  &=&-\frac12  \cD_+ \Phi \bar \cD_+\bar\Phi  
~, \\
\cT_{----}  &=& -\p_{--}\Phi \p_{--}\bar\Phi
~.
\label{Tfree}
 \eeqn
\esubeq
 This is indeed the correct supercurrent $\cR$-multiplet for a massless free theory. 
 
 \subsubsection{\texorpdfstring{$T\bar T$}{TTbar}-deformed action}
 Now, we switch to our $T\bar T$-deformed action \eqref{eq:susyTTbaraction}.  
 In terms of the notation employed in this appendix it is   given by
   \be
 S_\a= \frac14  \int \rd^2\sigma \,\rd\theta^+ \rd\bar\theta^+  \Big(- \ri\bar\Phi \p_{--} \Phi + \alpha 
 \frac{ \cD_+ \Phi  \bar\cD_+ \bar \Phi \p_\mm\Phi   \p_\mm \bar\Phi }
 {  1+\alpha  \mathcal X+\sqrt{1+2 \alpha   \mathcal X +\alpha^2 \mathcal Y^2}  }  \Big)
 ~.
 \ee
  Covariantising it, we get  
    \be
 S_\a=  \frac14 \int \rd^2\sigma\, \rd\theta^+ \rd\bar\theta^+  E^{-1} \Big(- \ri \bar{\hat \Phi } \nabla_{--}  \hat\Phi + \alpha 
 \frac{  \nabla_+\hat \Phi  \bar{ \nabla}_+ \bar{\hat \Phi} \nabla_\mm\hat\Phi    \nabla_\mm \bar{\hat\Phi }}
 {  1+\alpha  \mathcal{ \hat X}+\sqrt{1+2 \alpha   \mathcal {\hat X} +\alpha^2 \mathcal {\hat Y}^2}  }  \Big)
 ~.
 \label{covariantTTbar}
 \ee
where
\be
\mathcal { \hat X}= \nabla_\pp \hat\Phi  \nabla_\mm \bar{\hat \Phi }+ \nabla_\pp  \bar{\hat \Phi}  \nabla_\mm \hat \Phi 
~, \qquad 
\mathcal { \hat Y}= \nabla_\pp \hat\Phi  \nabla_\mm \bar{\hat \Phi }- \nabla_\pp  \bar{\hat \Phi}  \nabla_\mm \hat \Phi   
~.
 \ee

 Since the free part has been computed, we now focus on the second term, the non-linear part. 
 The three currents are considered separately.

 \vspace{0.1cm}
  \subsection*{$ \bullet$ Computation of $\cR_{++}$}
 \vspace{0.1cm}
 
The simplest current is $\cR_{++}$ 
which can be computed by  turning on only  $H_{--}{}^{++}$
while setting the other prepotentials to zero $H_{--}=H_{--}{}^{--}=0$.
 In this case, to leading order, the covariant derivatives  are given by
\bsubeq
 \beqn
   \nabla_{++ }&=&  \p_{++}   
  ~,~~~~~~
   \nabla_{+}=\cD_+
   ~,~~~~~~
    \bar \nabla_{+}=\bar\cD_+  ~,
    \\
 \nabla_{--}&=& \p_{--} -H_{--}{}^{++} \p_{++} 
 +\frac{\ri}{2} \bar \cD_+ H_{--}{}^{++} \cD_+ -  \frac{\ri}{2}  \cD_+ H_{--}{}^{++} \bar\cD_+ 
 ~.
 \eeqn
\esubeq
The superdensity is $E^{-1}=1$ and $\hat \Phi =\Phi, \bar{\hat \Phi}=\bar\Phi$. 
 The numerator of the nonlinear part  is
 \be
  \nabla_+\hat \Phi  \bar{ \nabla}_+ \bar{\hat \Phi} =\cD_+\Phi_+ \bar\cD_+\bar\Phi_+
  ~.
 \ee
 Due to the fermionic nature of this term,  the denominators of  the nonlinear part can not have terms like 
 $\cD_+\Phi_+, \bar\cD_+\bar\Phi_+$. So \text{eff}ectively, we can use 
 \be
  \nabla_{--} = \p_{--} -H_{--}{}^{++} \p_{++} 
  ~.
 \ee
Ultimately,  it is easy to find, to linear order in the prepotentials
 \bea
  \frac{   \nabla_\mm\hat\Phi    \nabla_\mm \bar{\hat\Phi }}
 {  1+\alpha  \mathcal{ \hat X}+\sqrt{1+2 \alpha   \mathcal {\hat X} +\alpha^2 \mathcal {\hat Y}^2}  }
& =& \frac{  \p_\mm\Phi   \p_\mm \bar\Phi }
 {  1+\alpha  \mathcal X+\sqrt{1+2 \alpha   \mathcal X +\alpha^2 \mathcal Y^2}  }  
 \non\\
 &&
 + \frac{H_{--}{}^{++}}{2\alpha}  \Big( \frac{1}{\sqrt{1+2 \alpha   \mathcal X +\alpha^2 \mathcal Y^2}  }  -1\Big)
~.
 \eea
 Then, it is straightforward to extract the $\cR_{++}$ supercurrent which takes the form
\be
 \cR_{++}=-  \frac{\cD_+\Phi \bar\cD_+\bar\Phi}{ 2 \sqrt{1+2 \alpha   \mathcal X +\alpha^2 \mathcal Y^2}  } 
 ~.
 \label{R++}
\ee 
 where the free part contribution has also been included. 
 
  \vspace{0.1cm}
  \subsection*{$ \bullet$ Computation of $\cR_{-- }$}
  \vspace{0.1cm}

 Now we set $H_{--}=H_{--}{}^{++}=0$ and keep only  $H_{--}{}^{--}$. Then, the covariant derivatives  become 
\bsubeq
 \beqn
 \nabla_+ &=&\cD_+  -H_+{}^+ \cD_+
 ~,\\
 \bar \nabla_+ &=&\bar \cD_+  -H_{\bar +}{}^{\bar +} \bar\cD_+
 ~.
 \eeqn
\esubeq
 
The  superdensity is $E^{-1}=(1+H_{--}{}^{--})$ and $\hat \Phi =\Phi, \bar{\hat \Phi}=\bar\Phi$. 
 In this case 
  the numerator of \eqref{covariantTTbar}, at the linear order, takes the form
  \be
  \nabla_+\hat \Phi  \bar{ \nabla}_+ \bar{\hat \Phi} 
=
  (1-H_{--}{}^{--})\cD_+\Phi_+ \bar\cD_+\bar\Phi_+
  ~.
 \ee
 
  Using similar arguments as   in the last subsection, effectively we have 
\bsubeq
 \beqn
  \nabla_{--} &=& \p_{--} -H_{--}{}^{--} \p_{--}  
  ~, \\
  \nabla_{++} &=& \p_{++} -H_{--}{}^{--} \p_{++}   
  ~.
 \eeqn
\esubeq 
 
 Plugging these results back into the action and expanding to linear order in the prepotentials, one finds the supercurrent 
 \be
\cR_{--}=  \alpha \frac{  \cD_+ \Phi  \bar \cD_+ \bar \Phi   }{   \sqrt{1+2 \alpha   \mathcal X +\alpha^2 \mathcal Y^2} } 
 \frac{  \p_\mm \bar \Phi  \p_\mm  \Phi }{1+\alpha   \mathcal X+\sqrt{1+2 \alpha   \mathcal X +\alpha^2 \mathcal Y^2}}
 ~.
 \ee

   \vspace{0.1cm}
 \subsection*{$ \bullet$ Computation of $\cT_{---- }$}
  \vspace{0.1cm}
  
  Finally, we are going to calculate $\cT_{---- }$  by just turning on $H^{--}$ and by setting $H_{--}{}^{--}=H_{--}{}^{++}=0$.
   The covariant derivatives   are 
\bsubeq
 \beqn
 \nabla_+ &=& \cD_+ -\ri\cD_+ H^{--} \p_{--} 
 ~,\\
 \bar\nabla_{  +} &=& \bar  \cD_+ +\ri  \bar \cD_+ H^{--} \p_{--} 
 ~,\\
  \nabla_{++} &=& \p_{++} +\frac12 [\cD_+, \bar\cD_+] H^{--}\p_{--} 
  ~,\\
    \nabla_{--} &=& \p_{--}
    ~.
 \eeqn
 \esubeq
 The  superdensity is simply given by $E^{-1}=1$, but the convariantly chiral superfield and its covariant derivative
  have a nontrivial dependence 
upon $H^{--}$:
 \be
 \hat \Phi=(1-\ri H^{--} \p_{--}) \Phi
 ~,~~~~~~~
  \nabla_+  \hat \Phi=(\cD_+ - 2\ri \cD_+ H^{--} \p_{--}) \Phi
  ~.
 \ee
 The extra pieces here poses an obstruction to further simplifying the covariant derivatives  as  we did before. 
 This makes the calculation of $\cT_{---- }$ more complicated. 
Collecting all the results together, we get the covariantised action 
expanded to first order in $H^{--}$
 \beqn
 S_{\text{int}} &=& 
\frac14 \int \rd^2\sigma \,\rd\theta^+ \rd\bar\theta^+\;\Bigg[ \alpha \cD_+\Phi \bar\cD_+ \bar\Phi \cdot [\cD_+, \bar \cD_+]H^{--} (-\alpha)(\p_\mm\Phi   \p_\mm \bar\Phi)^2  \frac{1+V}{V Z^2}
\non\\ &&
 +2\ri \alpha \Big(\cD_+\Phi \cdot\bar \cD_+ H^{--} \p_{--}\bar\Phi + \bar \cD_+ \bar \Phi \cdot  \cD_+ H^{--} \p_{--} \Phi     \Big)
 \frac{ \p_\mm\Phi   \p_\mm \bar\Phi}{Z} 
 \Bigg] 
 ~,
 \label{Tcomput}
 \eeqn
  where (we also introduce $\tilde Z$ for later convenience)
\bsubeq
 \bea
V&=&  \sqrt{1+2 \alpha   \mathcal X +\alpha^2 \mathcal Y^2}~, \\
  Z&=& 1+\alpha  \mathcal X+\sqrt{1+2 \alpha   \mathcal X +\alpha^2 \mathcal Y^2}~, \\
\tilde Z&=& 1+\alpha  \mathcal X-\sqrt{1+2 \alpha   \mathcal X +\alpha^2 \mathcal Y^2}~.
 \eea
 \esubeq
 
To simplify the calculation, from now on we  focus on  terms contributing to the supercurrent
which have no bare $\cD_+\Phi$ and $ \bar\cD_+ \bar\Phi $ terms. 
The reason is that, when we consider the $T\bar T$ primary operator, the contribution involving $\cT_{----}$
will appear in the  product $\cT_{----}  \cR_{++} $. 
From the explicit expression of $\cR_{++}$ \eqref{R++}, 
we immediately see that any $ \cD_+\Phi ,\bar\cD_+ \bar\Phi $  part  in $  \cT_{----} $ 
has no contributions due to its fermionc nature.
After some integration by parts in \eqref{Tcomput} we obtain 
\beqn
S_{\text{int}} &=& 
\frac14 \int \rd^2\sigma \,\rd\theta^+ \rd\bar\theta^+\
 2   H^{--}(\p_\mm\Phi   \p_\mm \bar\Phi)  \Big[
   \alpha^2( \mathcal X^2- \mathcal Y^2 )   \frac{1+V}{VZ^2}
-2 \alpha    \frac{ \mathcal X}{Z}
+ \cdots
\Big]
~,~~~~~~~~~
\non
\\&=&
\frac14 \int \rd^2\sigma \,\rd\theta^+ \rd\bar\theta^+\
 2   H^{--}(\p_\mm\Phi   \p_\mm \bar\Phi)  \Big[
  \frac{1 }{ V} -1
+ \cdots 
\Big]
~,
\eeqn
 where  the ellipsis represent terms  proportional to $\cD_+\Phi$ or  $\bar\cD_+ \bar\Phi$, and we   used the relations
\be
\alpha^2( \mathcal X^2- \mathcal  Y^2 ) =Z\tilde Z
~, \qquad 
\frac{1}{V}-\frac{\tilde Z}{V   Z}=\frac{2}{Z}
~.
\ee
Once we add the contribution form the free action \eqref{Tfree} we obtain
 \be
    \cT_{----} =- \frac{  \p_\mm\Phi   \p_\mm \bar\Phi }{  \sqrt{1+2 \alpha   \mathcal X +\alpha^2 \mathcal Y^2}}+  
     \cD_+\Phi \cdot \#  +  \bar\cD_+ \bar\Phi \cdot \#  
     ~.
 \ee

It is easy to verify that these supercurrents give rise to the correct energy-momentum tensor in the pure bosonic case.
 This also  enables us to translate the results here into  the notation of the main body of the paper.

\section{Deformation of the free action in components}\label{appComponent}
We have given schematically the component expression of the supersymmetric $T\bar{T}$ deformation of a free 
$\mathcal{N}=(0,2)$ action in eq.~\eqref{susy-S-components}. Here we give the explicit form of the coefficients,
 in terms of the variable $x,y$ introduced in eq.~\eqref{eq:xydef}.
\begin{equation}
\begin{aligned}
A(x,y)&= \frac{1}{4\alpha}\Big(-1+\sqrt{1+2\alpha x+ \alpha^2y^2}\Big)
\,,\\
B(x,y)&=  \frac{\ri}{4\sqrt{1+2\alpha x+\alpha^2 y^2}}
\,,\\
C(x,y)&=\frac{\ri}{2\alpha (x^2-y^2)}\Big(1-\frac{1+\alpha x}{\sqrt{1+2\alpha x + \alpha^2 y^2}}\Big)
\,,\\
D_{--}(x,y)&=\frac{\ri\alpha}{4\sqrt{1+2\alpha x+\alpha^2 y^2}}
\Bigg[(1+\alpha (x +2y))\Big(\partial_{--}\partial_{--} \phi \partial_{++}\bar{\phi}
+ \partial_{--}\partial_{++} \bar{\phi} \partial_{--}{\phi}\Big)
\Bigg]\\
&\quad+\ri\partial_{--}\phi(\partial_{--}\bar{\phi})^2\partial_{++}^2\phi
\Bigg[\frac{\sqrt{1+2\alpha x+\alpha^2y^2}}{\alpha (x-y)(x+y^2)}
+\frac{3}{2\alpha(x-y)^2(x+y)}\\
&\qquad\qquad\qquad\qquad\qquad\qquad
+\frac{1+3\alpha^2 x^2-\alpha^3(2x^3-x^2y-4xy^2+y^3)}{2\alpha(x-y)^2(x+y)(1+2\alpha x+\alpha^2y^2)}\Bigg]\\
&\quad +{\rm  c.c.}
\, ,\\
E(x,y)&=-\frac{\alpha}{4}\frac{1-\alpha x}{\big(1+2\alpha x +\alpha^2 y^2\big)^{3/2}}
\,,\\
F(x,y)&=\frac{\alpha^2}{\big(1+2\alpha x +\alpha^2 y^2\big)^{3/2}}
\,,\\
G(x,y)&=\frac{4}{\alpha(x^2-y^2)^2}\Big(1-\frac{(1+\alpha x)}{\sqrt{1+2\alpha x +\alpha^2y^2}}
+\frac{x^2-y^2}{2}\frac{\alpha^2}{\big(1+2\alpha x+\alpha^2y^2\big)^{3/2}}\Big)
\,.
\end{aligned}
\end{equation}


\bibliographystyle{JHEP}
\bibliography{refs}

\end{document}